\documentclass[twocolumn]{IEEEtran}
\usepackage{ifpdf}
\usepackage{cite}
\usepackage{bbm}
\ifCLASSINFOpdf
\usepackage{graphicx}
\usepackage{subfigure}
\usepackage{epstopdf}
\usepackage{xcolor}
\else
\fi
\usepackage{float}
\usepackage{amsmath}
\usepackage{amssymb}
\usepackage{amsthm}
\usepackage{mathtools}
\DeclareMathOperator*{\argmin}{arg\,min}

\usepackage{algorithm}
\usepackage{algpseudocode}

\allowdisplaybreaks
\usepackage{array}
\usepackage{fixltx2e}
\usepackage{stfloats}
\usepackage{tikz}
\theoremstyle{definition}

\usepackage[shortlabels]{enumitem}
\usepackage{comment}
\usepackage{multirow}
\usepackage{cleveref}

\begin{document}
\title{Universal Joint Source-Channel Coding for Modulation-Agnostic Semantic Communication}


\author{
	\IEEEauthorblockN{Yoon Huh},~\IEEEmembership{Graduate Student Member,~IEEE},
    \IEEEauthorblockN{Hyowoon Seo},~\IEEEmembership{Member,~IEEE},
	and
	\IEEEauthorblockN{Wan Choi},~\IEEEmembership{Fellow,~IEEE}
    \thanks{This work was supported by Institute of Information \& communications Technology Planning \& Evaluation (IITP) grant funded by the Korea government(MSIT) (No.RS-2024-00398948, Next Generation Semantic Communication Network Research Center)}
	\thanks{Y.~Huh and W.~Choi are with the Departement of Electrical and Computer Engineering, and the Institute of New Media and Communications, Seoul National University (SNU), Seoul 08826, Korea (e-mail: \{mnihy621, wanchoi\}@snu.ac.kr)}
    \thanks{H.~Seo is with the Department of Electrical and Computer Engineering, Sungkyunkwan University, Suwon 16419, South Korea (e-mail: hyowoonseo@skku.edu)}
    \thanks{(\emph{Corresponding Authors: Hyowoon Seo and Wan Choi})}
        \vspace{-3mm}

}

%

\maketitle

\begin{abstract}
From the perspective of joint source-channel coding (JSCC), there has been significant research on utilizing semantic communication, which inherently possesses analog characteristics, within digital device environments. However, a single-model approach that operates modulation-agnostically across various digital modulation orders has not yet been established. This article presents the first attempt at such an approach by proposing a universal joint source-channel coding (uJSCC) system that utilizes a single-model encoder-decoder pair and trained vector quantization (VQ) codebooks. To support various modulation orders within a single model, the operation of every neural network (NN)-based module in the uJSCC system requires the selection of modulation orders according to signal-to-noise ratio (SNR) boundaries. To address the challenge of unequal output statistics from shared parameters across NN layers, we integrate multiple batch normalization (BN) layers, selected based on modulation order, after each NN layer. This integration occurs with minimal impact on the overall model size. Through a comprehensive series of experiments, we validate that the modulation-agnostic semantic communication framework demonstrates superiority over existing digital semantic communication approaches in terms of model complexity, communication efficiency, and task effectiveness.
\end{abstract}

\begin{IEEEkeywords}
semantic communication, modulation-agnostic, joint source-channel coding, vector quantization, batch normalization.
\end{IEEEkeywords}

\IEEEpeerreviewmaketitle

\section{Introduction}
We now have access to intelligent services encompassing smart X (e.g., home, city, vehicle factory, healthcare, etc.) and extended (augmented and virtual) reality, previously confined to the realm of imagination. However, the widespread adoption of such services faces a significant obstacle from the perspective of communication engineering. This challenge pertains to meeting user demands, often referred to as quality of experience (QoE), by efficiently and swiftly transmitting vast amounts of data. In response, sixth-generation (6G) communication has emerged to address these demands, heralding an anticipated paradigm shift from data-oriented to service (or task)-oriented communication \cite{shi2023task}.

A pivotal driver of this transition is semantic communication \cite{gunduz2022beyond, seo2023semantics, Seo2024}, evolving in tandem with advancements in machine learning (ML). Semantic communication targets the resolution of Level B (semantic) and Level C (effectiveness) communication issues \cite{weaver1953recent, luo2022semantic}, with task-oriented joint source-channel coding (JSCC) at its core \cite{farsad2018deep, bourtsoulatze2019deep, xie2021deep, xu2021wireless}. Unlike traditional communication systems, which segregate source coding and channel coding modules, often with some loss of optimality, semantic communication, geared towards enhancing task performance, underscores an undeniable optimality gap in such modular separation. Consequently, current research in semantic communication focuses on harnessing ML capabilities to jointly optimize task execution, source coding, and channel coding, culminating in the implementation of neural network (NN)-based JSCC encoder-decoder pairs.

The challenge lies in the nature of NN training reliant on backpropagation, necessitating the exploration of JSCC techniques to facilitate the exchange of feature vectors within a continuous vector space between encoders and decoders. Although the implementation of semantic communication through analog transceivers might appear straightforward, it faces significant compatibility issues with the prevailing digital device infrastructure. An alternative strategy involves encoding the entries of feature vectors at the bit level and utilizing traditional digital modulation techniques. However, this approach is often inefficient, diminishing the benefits associated with semantic communication. Therefore, for effective semantic communication within a digital context, it is essential to incorporate digital modulation into the JSCC design considerations.

\begin{figure}[!t]
    \centering
    \includegraphics[width=\linewidth]{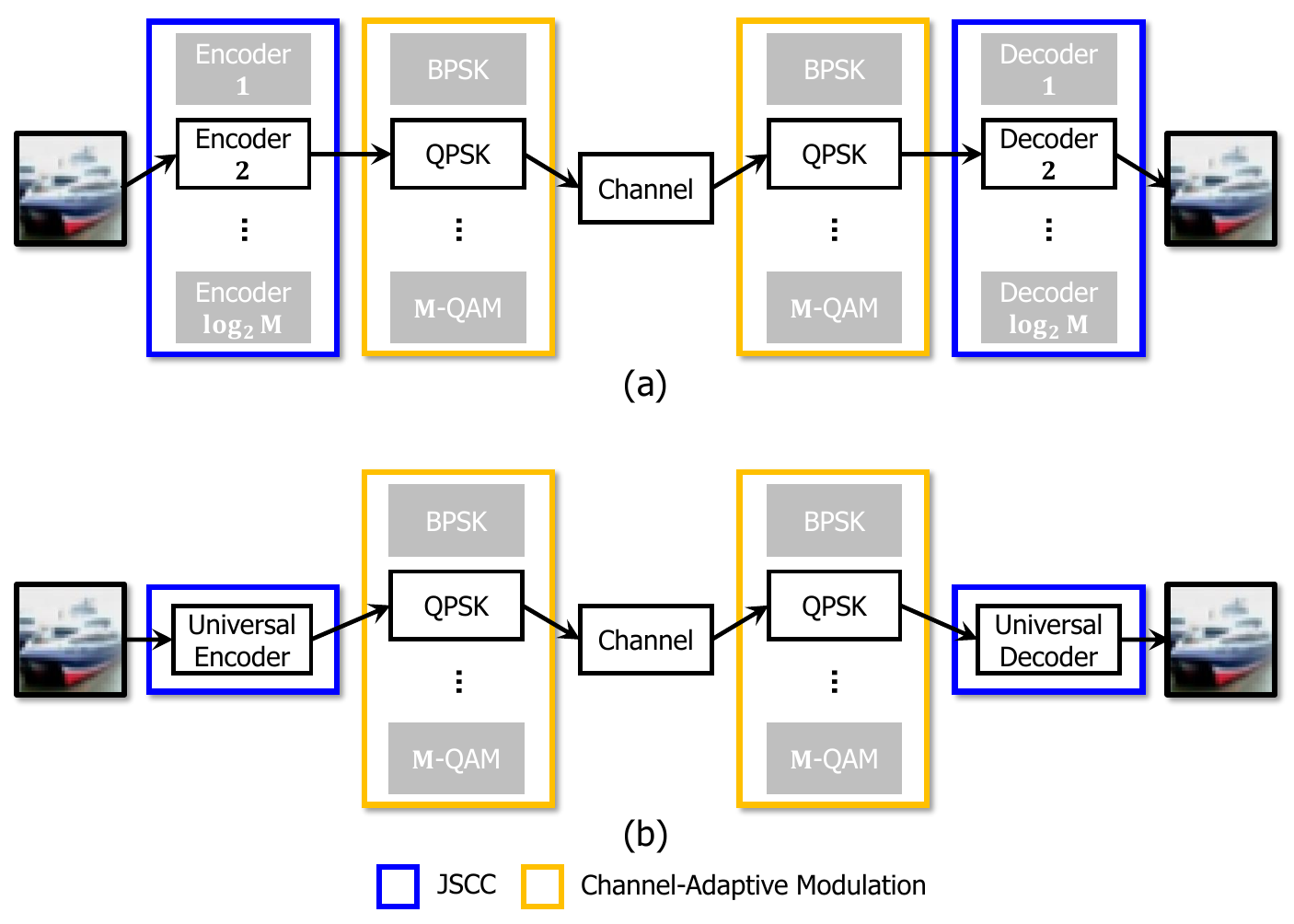}
    \vspace{-7mm}
    \caption{An illustration of (a) dedicated JSCC encoders/decoders and (b) universal encoder/decoder for semantic communication with channel-adaptive modulation.}
    \label{fig:Introduction}
    \vspace{-5mm}
\end{figure}

In the contemporary research landscape, innovative methods have been developed, notably the application of vector quantization (VQ) to the output feature vectors of encoders, followed by conventional digital modulation schemes (e.g., BPSK, 4QAM, 16QAM) or machine learning (ML)-based modulation utilizing a trained constellation map. These approaches, cited in recent studies \cite{hu2023robust, xie2023robust, tung2022deepjscc}, demonstrate robust performance within a specific modulation order. Yet, the goal of achieving consistent semantic communication performance across various modulation orders--—a concept termed \emph{modulation-agnostic semantic communication}\footnote{Throughout the article, the modulation techniques discussed all pertain to digital modulation, and modulation-agnostic semantic communication refers to a digital semantic communication system designed to transmit or receive information without being constrained by any specific modulation scheme.}—--requires a comprehensive suite of JSCC encoder-decoder pairs, as depicted in the blue boxes of Fig. \ref{fig:Introduction}(a). This approach can certainly optimize the task performance under varying channel conditions, which is a core objective of consistent semantic communication, as the optimal modulation order is typically dependent on the channel environment. However, maintaining such flexibility is often inefficient in terms of learning and hardware costs, including memory usage. Consequently, the pursuit of a universal framework capable of operating seamlessly across different modulation orders remains crucial for enhancing efficiency.

In response to the outlined challenges, as illustrated in Fig. \ref{fig:Introduction}(b), this article introduces a novel framework for modulation-agnostic semantic communication, which is anchored in a single-model universal JSCC scheme that is adept at functioning across various modulation orders. To realize this, an NN-based uJSCC encoder-decoder pair in conjunction with a VQ codebook is developed to utilize conventional digital modulation schemes. Notably, our proposed single-model encoder-decoder system demonstrates effective communication and task performance while maintaining reduced model complexity and training overhead compared to other benchmark approaches. In addition, the VQ codebook is not uniformly designed but is trained jointly with the uJSCC encoder-decoder, ensuring robust transmission of the feature vectors.

The proposed framework is validated through a series of experiments in a semantic communication setting for image transmission and reconstruction tasks. We compare the proposed uJSCC against benchmark schemes, including model-efficient JSCC (which leverages a single model trained simultaneously for all modulation orders) and task-effective JSCC (which independently trains multiple models for each modulation order). The proposed scheme outperforms the model-efficient approach and achieves comparable or superior task performance compared to the task-effective JSCC by employing multiple batch normalization (BN) modules at each CNN layer. The effectiveness of multiple BN modules in this context is clearly verified through ablation studies. Additionally, DeepJSCC-Q \cite{tung2022deepjscc} and conventional separate source-channel coding (SSCC) methods are employed as benchmark schemes. Furthermore, experiments demonstrate that the proposed method consistently delivers excellent performance across different encoder-decoder model sizes, with a higher volume of symbol transmission, and for various datasets including low and high-resolution images.

\subsection{Related works}

\subsubsection{Digital Semantic Communication}
A recent study \cite{hu2023robust} introduced a deterministic encoder-decoder pair in a digital semantic communication system using a VQ-variational autoencoder (VQ-VAE) \cite{van2017neural} trained with a straight-through estimator and a vision transformer \cite{dosovitskiy2020image} backbone. The system enhances robustness against semantic noise through adversarial training and selective masking of non-critical image patches, identified via a feature importance module. However, there is a mismatch between codebook cardinality and modulation order, requiring two 16QAM symbols to transmit one index from a 256-cardinality codebook.

In \cite{xie2023robust}, the authors proposed a VQ-based JSCC system employing digital modulation and an information bottleneck (IB) \cite{tishby2000information} loss function. The loss function's hyperparameter regulates the balance between task-relevant information and coded redundancy, thereby adjusting codebook robustness. During training, they utilized Gumbel-softmax \cite{jang2016categorical}, using the inner product between feature vectors and codewords as logits. This results in stochastic encoding, where codeword indices are sampled based on logits. However, despite the IB-optimized number of encoded bits, the actual transmission did not reflect it because both codebook cardinality and modulation order were fixed.
Similarly, the encoder in \cite{tung2022deepjscc} generates complex values, with symbols sampled based on logits from distances between complex output values and constellation map symbols. Another approach in \cite{bo2024joint} involved a stochastic encoder trained with a mutual information-based loss function, also utilizing Gumbel-softmax during training. In this system, symbols are directly sampled from output logits, eliminating the need for a learned codebook. However, these studies \cite{xie2023robust, bo2024joint, tung2022deepjscc} face limitations in adapting to multiple modulation orders as they require different neural network-based encoder-decoder parameter pairs for each order, just as illustrated in Fig. \ref{fig:Introduction}(a).

\subsubsection{Modulation-Agnostic Semantic Communication}
In recent work \cite{gao2023adaptive}, the optimization of modulation order was addressed by solving a robustness verification problem to achieve a certain robustness probability, following the training of a semantic encoder and decoder. Another study \cite{he2023rate} optimized the channel coding rate for conventional LDPC channel code through a similar robustness verification problem. This work utilized the CROWN method \cite{zhang2018efficient} to assess feature importance and measure output perturbation gaps from pretrained semantic encoders and decoders. Since a separate source-channel coding system was employed, modulation order was aligned with the coding rate. However, both \cite{gao2023adaptive} and \cite{he2023rate} did not incorporate digital modulation in the training of their encoders and decoders; instead, they applied fixed-bit quantization to analog outputs from the semantic encoder during inference to facilitate digital modulation, indicating a lack of joint design with the digital communication framework.

Meanwhile, \cite{park2024joint} describes a system where a stochastic encoder produces logits to sample bits, thereby creating bit sequences. The authors used a binary symmetric erasure channel in their training scenario, simplifying the process by assuming a 4QAM context. Following training, the modulation order is adjusted according to channel SNR, with thresholds established based on predetermined robustness levels for each bit position. These levels are taken into account during training as well. However, the rationale behind training under 4QAM conditions was not clearly articulated, and the determination of robustness levels and SNR threshold adjustments was heuristic in nature, with the goal of achieving a smooth performance curve at the cost of increased complexity in system design.

\subsection{Contributions}
The contributions of this article are summarized as:
\begin{itemize}
    \item This study explores methodologies that enable semantic communication within digital communication environments. Notably, it proposes a modulation-agnostic semantic communication framework that incorporates all the digital modulation orders in training. 
    \item The cornerstone of the proposed framework is a uJSCC scheme that operates based on a trained encoder-decoder pair and a VQ codebook. The uJSCC scheme is capable of maintaining semantic communication as the optimal modulation order adjusts to varying channel conditions. To the best of our knowledge, this article presents the first instance of using a uJSCC scheme as a universal model capable of digitalizing semantic communication across multiple modulation orders.
    \item Through extensive experimentation, we demonstrate that the proposed uJSCC surpasses other benchmark schemes--—including model-efficient and task-effective approaches—--in terms of model and training complexity, as well as task performance.
    \item The numerical results illustrate that the proposed model achieves satisfactory reconstruction quality across all signal-to-noise ratio (SNR) ranges on both low-resolution and high-resolution datasets, utilizing an efficiently trained universal NN structure. Additionally, the analysis confirms that uJSCC can be optimized for larger model sizes and more transmitted symbols, demonstrating its generality with fewer parameters.
\end{itemize}

\subsection{Notations}
Vectors and matrices are expressed in lower case and upper case bold, respectively. $\|\cdot\|_2$ denotes the $l_2$ norm of a vector. $\mathbb{R}$ and $\mathbb{Z}$ represent, respectively, the real number set and the integer set. $[a{:}b]$ is a integer set of $\{a, a + 1, \dots, b\}$. $[\cdot]^T$ denotes the transpose of a matrix or vector. $[\mathbf{A}]_{i:j, k:l}$ and $[\mathbf{a}]_{i:j}$ represent, respectively, the sliced matrix of $\mathbf{A}$ with row indices from $i$ to $j$ and column indices from $k$ to $l$, and the sliced vector of $\mathbf{a}$ with indices from $i$ to $j$. $\mathcal{CN}(\boldsymbol{\mu}, \boldsymbol{\Sigma})$ is complex normal distribution with mean vector $\boldsymbol{\mu}$ and covariance matrix $\boldsymbol{\Sigma}$. $\mathbf{I}_m$ is $m\times m$ identity matrix. $\mathbf{E}[\cdot]$ denotes the expected value of a given random variable.

\section{Preliminaries}\label{sec:Preliminaries}

\subsection{System Description}\label{subsec:System Description}

\begin{figure}[!t]
    \centering
    \includegraphics[width=\linewidth]{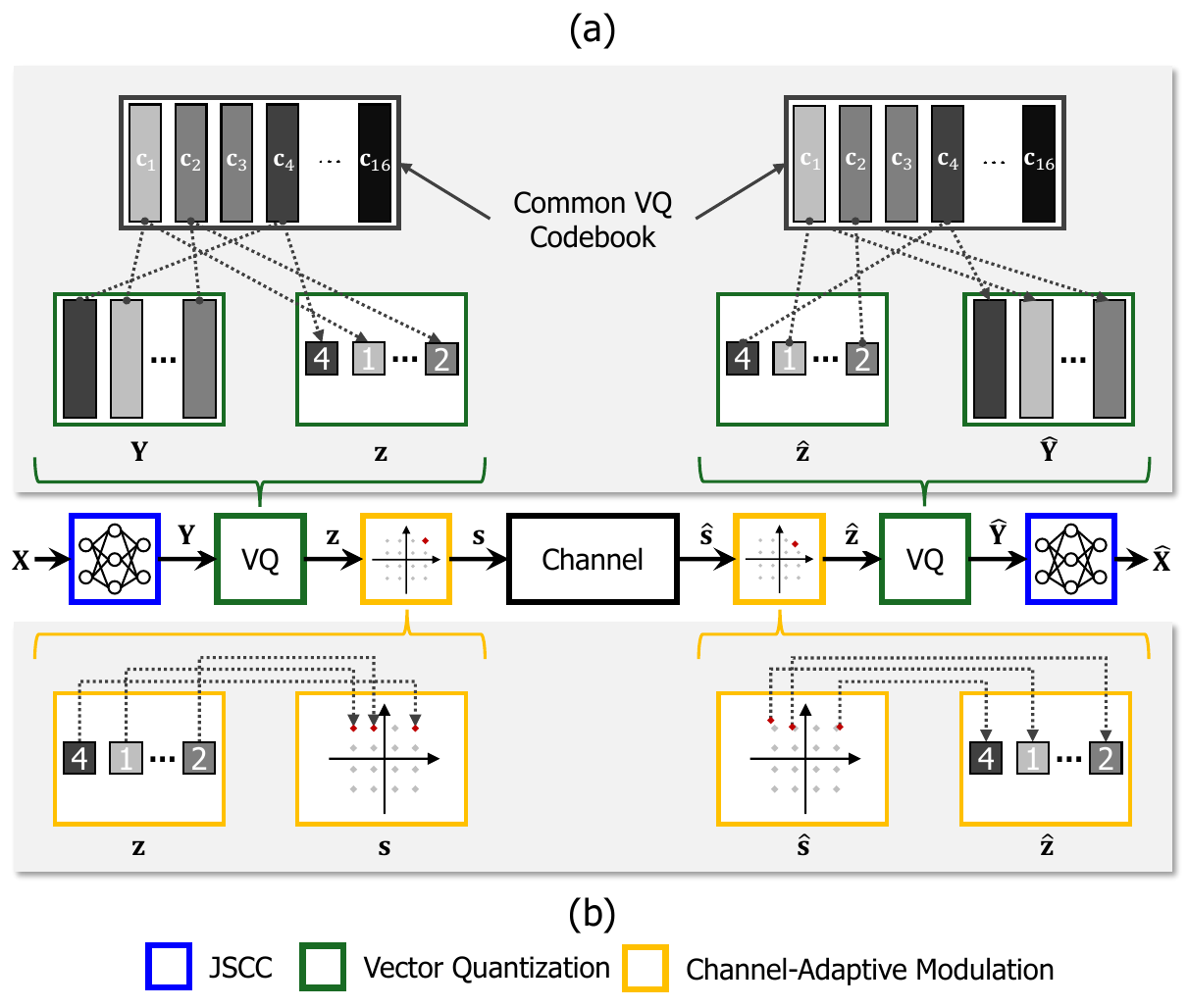}
    \vspace{-7mm}
    \caption{An illustration of (a) VQ codebook utilization and (b) constellation symbol mapping with 16QAM ($m = 16$).}
    \label{fig:System model}
    \vspace{-5mm}
\end{figure}

Consider a point-to-point semantic communication system designed for image transmission tasks, incorporating JSCC encoder-decoder, vector quantizer-dequantizer, and modulator-demodulator pairs, as depicted in Fig. \ref{fig:System model}. For clarity and initial exposition, we assume a fixed digital modulation scheme characterized by order $m$. However, we subsequently investigate the system's adaptability to accommodate a spectrum of modulation-order options. The operational structure and functionality of the system are described in full below.

The transmitter encodes a source image $\mathbf{X}\in\mathbb{R}^{C\times H\times W}$, where $C$, $H$, and $W$ are the number of channels, the height and width of the source image, respectively, into $N$ feature vectors of $D$ dimensions $\mathbf{Y} = [\mathbf{y}_1, \mathbf{y}_2, \dots, \mathbf{y}_N]^T\in\mathbb{R}^{N\times D}$ via an encoder $f_{\boldsymbol{\theta}}$ parameterized by $\boldsymbol{\theta}$, such that $\mathbf{Y} = f_{\boldsymbol{\theta}}(\mathbf{X})$. Then, as depicted in Fig. \ref{fig:System model}(a), VQ is done utilizing a codebook with $m$ codewords $\mathbf{C} = [\mathbf{c}_1, \mathbf{c}_2, \dots, \mathbf{c}_m]^T\in\mathbb{R}^{m\times D}$ with the minimum distance criterion, i.e., the codeword index $z_i$ for the $i$-th feature vector $\mathbf{y}_i$ is chosen as $z_i = \argmin_{j} \|\mathbf{y}_i - \mathbf{c}_j\|_2^2$ for all $i\in[1{:}N]$.

As illustrated in Fig. \ref{fig:System model}(b), each codeword is mapped one-to-one to a symbol on a given modulation constellation map $\mathbb{S} = \{\mathbf{s}_1, \mathbf{s}_2, \dots, \mathbf{s}_m\}$, where $\mathbf{s}_i\in\mathbb{R}^2$ and $\mathbf{E}[\|\mathbf{s}_i\|_2^2] = P$ for all $i\in[1{:}m]$, e.g., if $z_2 = 1$, its corresponding constellation symbol is $\mathbf{s}_1$. The index vector $\mathbf{z} = [z_1, z_2, \dots, z_N]^T\in[1{:}m]^N$ is modulated into a symbol sequence $\mathbf{s} = [\mathbf{s}_{z_1}, \mathbf{s}_{z_2}, \dots, \mathbf{s}_{z_N}]\in\mathbb{S}^N$ and sent through an additive white Gaussian noise (AWGN) channel. At the receiver side, the receiver gets the symbol sequence with AWGN noise $\mathbf{n} \sim \mathcal{CN}(0, \sigma^2\mathbf{I}_N)$ as $\hat{\mathbf{s}} = \mathbf{s} + \mathbf{n}$, where SNR is calculated as $\frac{P}{\sigma^2}$. Then the receiver detects symbol sequence $\Bar{\mathbf{s}} = [\mathbf{s}_{\hat{z}_1}, \mathbf{s}_{\hat{z}_2}, \dots, \mathbf{s}_{\hat{z}_N}]\in\mathbb{S}^N$, and dequantization is done from the detected index vector $\hat{\mathbf{z}} = [\hat{z}_1, \hat{z}_2, \dots, \hat{z}_N]^T\in[1{:}m]^N$ utilizing the same codebook $\mathbf{C}$ as the transmitter side, and then results in $\hat{\mathbf{Y}} = [\mathbf{c}_{\hat{z}_1}, \mathbf{c}_{\hat{z}_2}, \dots, \mathbf{c}_{\hat{z}_N}]^T\in\mathbb{R}^{N\times D}$. Finally, the receiver can reconstruct the image $\hat{\mathbf{X}}\in\mathbb{R}^{C\times H\times W}$ from $\hat{\mathbf{Y}}$ via a decoder $g_{\boldsymbol{\phi}}$, parameterized by $\boldsymbol{\phi}$, as $\hat{\mathbf{X}} = g_{\boldsymbol{\phi}}(\hat{\mathbf{Y}})$.

To optimize the system's operation, it is essential to train the encoder parameters $\boldsymbol{\theta}$, decoder parameters $\boldsymbol{\phi}$, and the codebook $\mathbf{C}$. The optimization is guided by the loss function proposed in \cite{van2017neural}, which encompasses three terms:
\begin{align}
    \mathcal{L} = \mathbf{E}_{\mathbf{X}}[ \mathsf{e}(\hat{\mathbf{X}}, \mathbf{X}) + \alpha \mathsf{e}(\hat{\mathbf{Y}}, \mathbf{Y}_\mathsf{d}) + \beta \mathsf{e}(\mathbf{Y}, \hat{\mathbf{Y}}_\mathsf{d})],
\end{align}
where $\mathsf{e}(x, y)$ denotes the mean squared error between $x$ and $y$, and $\mathbf{A}_\mathsf{d}$ represents duplicated versions of $\mathbf{A}$. The duplication is implemented to inhibit the flow of gradients through $\mathbf{A}$, thereby preventing updates to $\mathbf{A}$ or the NN parameters involved in computing the matrix. The parameters $\alpha$ and $\beta$ serve as hyperparameters that regulate the relative significance of the respective loss components. The role of each component is summarized as follows.
\begin{itemize}
    \item The first term pertains to the quality of the reconstruction task, albeit hindered by the inability to apply the chain rule in gradient calculation due to VQ. Consequently, leveraging the straight-through gradient estimator between $\mathbf{Y}$ and $\hat{\mathbf{Y}}$, we establish a linkage between the gradient from $\mathsf{e}(\hat{\mathbf{X}}, \mathbf{X})$ to $\hat{\mathbf{Y}}$ and the gradient from $\mathbf{Y}$ to $\mathbf{X}$. This term significantly influences the training of both encoder parameter $\boldsymbol{\theta}$ and decoder parameter $\boldsymbol{\phi}$.
    \item The second term aims to minimize the disparity between the detected codewords $\hat{\mathbf{Y}}$ and the duplicated original feature vectors $\mathbf{Y}_\mathsf{d}$, thereby guiding the training of the codebook $\mathbf{C}$ alone.
    \item Due to challenges in achieving convergence of the codebook, the third term serves to regularize the training of the codebook $\mathbf{C}$, which is not directly related to the update of the codebook but the JSCC encoder. It achieves this by minimizing the disparity between the original feature vectors $\mathbf{Y}$ and the duplicated detected codewords $\hat{\mathbf{Y}}_\mathsf{d}$, thereby facilitating the alignment of the original feature vectors with the nearest codewords in the codebook.
\end{itemize}
Consequently, the parameters are obtained as $(\boldsymbol{\theta}^*, \boldsymbol{\phi}^*, \mathbf{C}^*) = \argmin_{\boldsymbol{\theta}, \boldsymbol{\phi}, \mathbf{C}} \mathcal{L}$. During the training process, it is crucial to manage the hyperparameter $\alpha$ with precision, while $\beta$ usually works well setting to $0.25\alpha$\cite{van2017neural}. If the sum of the second and third terms approaches zero too rapidly, the decoder will not receive sufficient training due to limited variation in the input codewords. In this scenario, reducing $\alpha$ is advisable. Conversely, if the sum of the second and third terms remains excessively large, increasing $\alpha$ is recommended since it indicates a substantial vector quantization error in the current codebook.

\subsection{Problem Formulation}\label{subsec:Problem Formulation}
In classical digital communication, a desired bit error rate (BER) or data rate is attained by adjusting the coding scheme or modulation order according to the channel conditions. Similarly, semantic communication could be envisioned to achieve optimal task performance by selecting the appropriate coding scheme or modulation order in response to the channel state.

Hence, the main target of the article is to develop a modulation-agnostic semantic communication framework, incorporating uJSCC encoder, decoder, and VQ codebook that support $K$ different modulation orders, i.e., $m_1, m_2, \dots m_K$, where $m_k$ represents the $k$-th modulation order and without loss of generality, $m_1 < m_2 < \dots < m_K$. In mathematical expression, let $\mathbf{B} = [b_1, b_2, \dots, b_{K - 1}]$ be the SNR boundary to choose modulation order among $K$ types, where
\begin{align}\label{eq:modulation order selection}
    m =
    \begin{cases}
        m_1, & \text{if}\quad \eta < b_1,\\
        m_k, & \text{if}\quad b_{k - 1} \leq \eta < b_{k},\\
        m_K, & \text{if}\quad \eta \geq b_{K - 1},
    \end{cases}
\end{align}
where $k\in[2{:}K - 1]$. The designing condition of the SNR boundary will be discussed in Section \ref{subsec:Universal Joint Source-Channel Coding}.

Assume that both transmitter and receiver are aware of the SNR value $\eta$, enabling synchronization of the modulation order between the transmitter and receiver. The following communication process is conducted with respect to modulation order index $k$ according to the SNR value $\eta$. A universal encoder parameterized by $\mathbf{\Theta}$, denoted by $\mathcal{F}_{\mathbf{\Theta}}$, encodes a source image $\mathbf{X}$ into feature vectors $\mathbf{Y}^{\scriptscriptstyle(k)}$ given the SNR value $b_{k-1} \leq \eta < b_k$, such that $\mathbf{Y}^{\scriptscriptstyle(k)} = \mathcal{F}_{\mathbf{\Theta}}(\mathbf{X}, \eta)\in\mathbb{R}^{N\times D_k}$ where $D_k$ is the dimension of feature vector of the modulation order $m_k$. The encoded feature vectors are quantized via codebook $\mathbf{C}^{\scriptscriptstyle(k)}$ and modulated into symbols $\mathbf{s}^{\scriptscriptstyle(k)}$. The codebook for modulation order $m_k$, which should also be trained, can be expressed as $\mathbf{C}^{\scriptscriptstyle(k)} = \left[\mathbf{c}_1^{\scriptscriptstyle(k)}, \mathbf{c}_2^{\scriptscriptstyle(k)}, \dots, \mathbf{c}_{m_k}^{\scriptscriptstyle(k)}\right]^T\in\mathbb{R}^{m_k\times D_k}$. At the receiver, the received symbols $\hat{\mathbf{s}}^{\scriptscriptstyle(k)}$ are detected and dequantized into $\hat{\mathbf{Y}}^{\scriptscriptstyle(k)}$ via the same codebook $\mathbf{C}^{\scriptscriptstyle(k)}$. Finally, a universal decoder parameterized by $\mathbf{\Phi}$, denoted as $\mathcal{G}_{\mathbf{\Phi}}$, decodes $\hat{\mathbf{Y}}^{\scriptscriptstyle(k)}$ into reconstructed image $\hat{\mathbf{X}}$ given $\eta$, such that $\hat{\mathbf{X}} = \mathcal{G}_{\mathbf{\Phi}}(\hat{\mathbf{Y}}^{\scriptscriptstyle(k)}, \eta)$.

The objective is to determine an optimal NN structure for the universal encoder $\mathcal{F}_{\mathbf{\Theta}}$ and universal decoder $\mathcal{G}_{\mathbf{\Phi}}$, which include data processing paths for $K$ modulation orders, codebook $\mathbf{C}^{\scriptscriptstyle(k)}$ for all $k \in [1{:}K]$, and the SNR boundary $\mathbf{B}$. When designing this system, considerations include model complexity, the total number of NN parameters, training strategies, and task performance.

\subsection{Intuitive Benchmark Schemes}\label{subsec:Benchmark Schemes}
Before delving into our proposed solution for the formulated problem above, we first introduce three benchmark schemes designed in a simplified manner to highlight the strengths and weaknesses of different approaches, aiding a deeper understanding of our solution. Note that additional benchmarks will be introduced in the experimental section for a more comprehensive evaluation.

\subsubsection{Model-Efficient JSCC}\label{subsubsec:ME}
The first and second benchmark schemes we introduce are what we refer to as the model-efficient (ME) JSCC. To minimize model complexity, this benchmark scheme utilizes a single encoder-decoder pair, but utilizes distinct VQ codebooks optimized for each modulation order. Given that higher modulation orders necessitate larger codeword dimensions, the codeword dimension $D_k$ is standardized to $D_K$ for all $k\in[1{:}K]$ within this benchmark, such that $\mathbf{Y}^{\scriptscriptstyle(k)}\in\mathbb{R}^{N\times D_K}$, contrary to the proposed uJSCC. This approach enables a uniform representation across various modulation orders, simplifying model management and potentially reducing both training and operational complexities. The model can be trained using two distinct types of training strategies, details of which will be elaborated upon in the following.

The first strategy, dubbed ME$_1$, consists of single-stage training. It jointly trains to optimize the image reconstruction performance of all modulation orders and obtains a single encoder-decoder pair and $K$ different codebooks for each modulation order. The loss function for the $k$-th modulation order can be represented as
\begin{align}\label{eq:mod loss function}
\begin{split}
    \mathcal{L}_k = \mathbf{E}_{\mathbf{X}}[\mathsf{e}(\hat{\mathbf{X}}^{\scriptscriptstyle(k)}, \mathbf{X}) &+ \alpha_k\mathsf{e}(\hat{\mathbf{Y}}^{\scriptscriptstyle(k)}, \mathbf{Y}^{\scriptscriptstyle(k)}_\mathsf{d})\\
    &+ \beta_k\mathsf{e}(\mathbf{Y}^{\scriptscriptstyle(k)}, \hat{\mathbf{Y}}^{\scriptscriptstyle(k)}_\mathsf{d})],
\end{split}
\end{align}
where the computation is done with the SNR value $\eta$ randomly sampled from the SNR range of the $k$-th modulation order. Then, the loss function to jointly train ME$_1$ can be expressed as $\mathcal{L}^{\scriptscriptstyle \mathrm{ME_1}} = \sum_{k = 1}^K\lambda_k\mathcal{L}_k$,
where $\lambda_k$ is a hyperparameter to control the importance of loss terms for each modulation order and $\mathcal{L}_k$ is calculated with a randomly sampled single $\eta$ value within the corresponding SNR range for one iteration.

The second strategy, termed ME$_2$, employs a double-stage training process. Initially, ME$_2$ focuses on optimizing the performance of semantic communication for the $K$-th modulation order, which is the highest in the sequence. The loss function for the initial stage is given by $\mathcal{L}^{\scriptscriptstyle \mathrm{ME_2}}_{\scriptscriptstyle (1)} = \mathbf{E}_{\eta}[\mathcal{L}_K]$.

Upon completion of the first stage, the second training stage commences. This stage is designed to optimize the codebooks $\mathbf{C}^{\scriptscriptstyle(k)}$ for the remaining modulation orders, encompassing all $k\in[1{:}K - 1]$, while keeping the encoder and decoder parameters fixed. The loss function for the second stage is defined as $\mathcal{L}^{\scriptscriptstyle \mathrm{ME_2}}_{\scriptscriptstyle (2)} = \sum_{k = 1}^{K - 1}\lambda_k\mathbf{E}_{\mathbf{X}}[\alpha_k\mathsf{e}(\hat{\mathbf{Y}}^{\scriptscriptstyle(k)}, \mathbf{Y}^{\scriptscriptstyle(k)}_\mathsf{d})]$.

\subsubsection{Task-Effective JSCC}\label{subsubsec:TE}
The third benchmark aims to optimize the performance of JSCC for all SNR ranges. To this end, a set of encoder-decoder pair and codebook for each modulation order $m_k$ should be trained independently with respect to $k$, different from ME approach and the proposed uJSCC. Then, during inference, a certain set for $m_k$ is switched to be used among $K$ sets, chosen according to the SNR value $\eta$ and $\mathbf{B}$. Such an approach is dubbed \emph{task-effective (TE)} approach. As the codebook cardinality for each modulation order increases with respect to $k$, we correspondingly scale the codeword dimensions $D_k$ to increase in length, such that $D_1 < D_2 < \dots < D_K$. The escalation in codeword dimension is strategically designed to ensure that the image reconstruction performance is not compromised by VQ errors.

\subsection{Discussions on Intuitive Benchmark Schemes}
The ME JSCC framework enables the accommodation of $K$ different modulation orders using a singular encoder-decoder pair, yielding a parameter count approximately $1/K$ times smaller than its TE JSCC counterpart. This substantial reduction in parameters simplifies the training process significantly. In terms of training complexity, ME JSCC methods streamline the procedure compared to TE JSCC. Specifically, the ME$_1$ variant optimizes NN parameters for $K$ modulation orders concurrently within a single epoch, whereas the TE strategy involves sequential training for each modulation order, focusing on individual encoder-decoder pairs. For ME$_2$, the initial training phase parallels the TE approach, targeting the $K$-th modulation order. Subsequently, the process becomes less complex by exclusively focusing on joint training of the codebooks for the remaining orders. This bifurcated strategy not only simplifies the training but also enhances efficiency by concentrating efforts on refining the most critical system elements.

Regarding task performance, TE JSCC is engineered to achieve superior performance across all SNR ranges. Conversely, the ME strategies exhibit inherent limitations and performance constraints. The ME$_1$ approach suffers from a performance ceiling due to parameter sharing across multiple modulation orders, ultimately constrained by TE capabilities. Moreover, with a uniform codeword dimension $D_k = D_K$ across all orders, the vector quantization error disproportionately affects lower modulation orders, degrading performance. In ME$_2$, although performance for the $K$-th modulation order may approach optimality due to targeted optimization, the lack of synergistic optimization between the encoder-decoder pair and the codebooks for other orders restricts performance across these lower orders.

To summarize, these solutions illustrate a compromise between model efficiency and task effectiveness within a modulation-agnostic semantic communication framework. This reveals a pressing need for an innovative, universally efficient system design that effectively balances satisfactory task performance with a suitable training algorithm.

\section{Modulation-Agnostic Universal Joint Source-Channel Coding}\label{sec:Modulation-Agnostic Universal Joint Source-Channel Coding}

This section presents a novel modulation-agnostic uJSCC scheme for digital semantic communication with its features summarized in TABLE \ref{tab:Features of uJSCC and intuitive benchmark schemes}. A primary challenge in uJSCC development is the shared use of CNN parameters across multiple data processing paths for each modulation order, which can lead to statistical discrepancies and degrade performance. We address this issue by implementing a switchable batch normalization layer. The training methodology for the proposed uJSCC system is also detailed.

\begin{table}[!t]
    \caption{Features of uJSCC and intuitive benchmark schemes}
    \label{tab:Features of uJSCC and intuitive benchmark schemes}
    \centering
    \begin{tabular}{|c|c|c|c|c|}
        \hline
        \textbf{Scheme} & \textbf{\# of Enc-Dec} & $\boldsymbol{D_k}$ & \textbf{Training} & \textbf{S-BN} \\ \hline
        \textbf{uJSCC}  & 1                      & Increasing         & Single            & O             \\ \hline
        \textbf{ME$_1$} & 1                      & $D_K$              & Single            & X             \\ \hline
        \textbf{ME$_2$} & 1                      & $D_K$              & Double            & X             \\ \hline
        \textbf{TE}     & $K$                    & Increasing         & Independent       & X             \\ \hline
    \end{tabular}
    \vspace{-5mm}
\end{table}

\subsection{Universal Joint Source-Channel Coding}\label{subsec:Universal Joint Source-Channel Coding}
Before proceeding, it is essential to discuss the methodology for setting the SNR boundaries mentioned previously. In accordance with the 3GPP technical specification \cite{3gpp.38.214}, the modulation and coding scheme (MCS) table delineates channel coding rates and modulation orders to maintain the block error rate (BLER) below 0.1. Our proposed system utilizes an ML strategy to train the channel coding scheme across various digital modulation methods and corresponding SNR ranges. Also note that, in the proposed system, one symbol corresponds to a single feature vector which contains much more information than information-theoretic bits in one symbol. Inspired by \cite{3gpp.38.214}, regarding a feature vector as the block in the conventional communication system, we establish SNR boundaries $\mathbf{B} \in \mathbb{Z}^{K - 1}$ aiming for a symbol error rate (SER) \cite{proakis2007fundamentals} around 0.1. This adjustment adapts conventional communication standards to the feature vectors in our system, facilitating joint training of the uJSCC encoder-decoder pair and enhancing robustness against data compression and channel distortion. For example, with modulation types such as BPSK, 4QAM, 16QAM, 64QAM, and 256QAM, where $K=5$, the SNR boundaries would be set at $b_1 = 5$ dB, $b_2 = 12$ dB, $b_3 = 20$ dB, and $b_4 = 26$ dB, respectively.

\begin{figure}[!t]
    \centering
    \includegraphics[width=\linewidth]{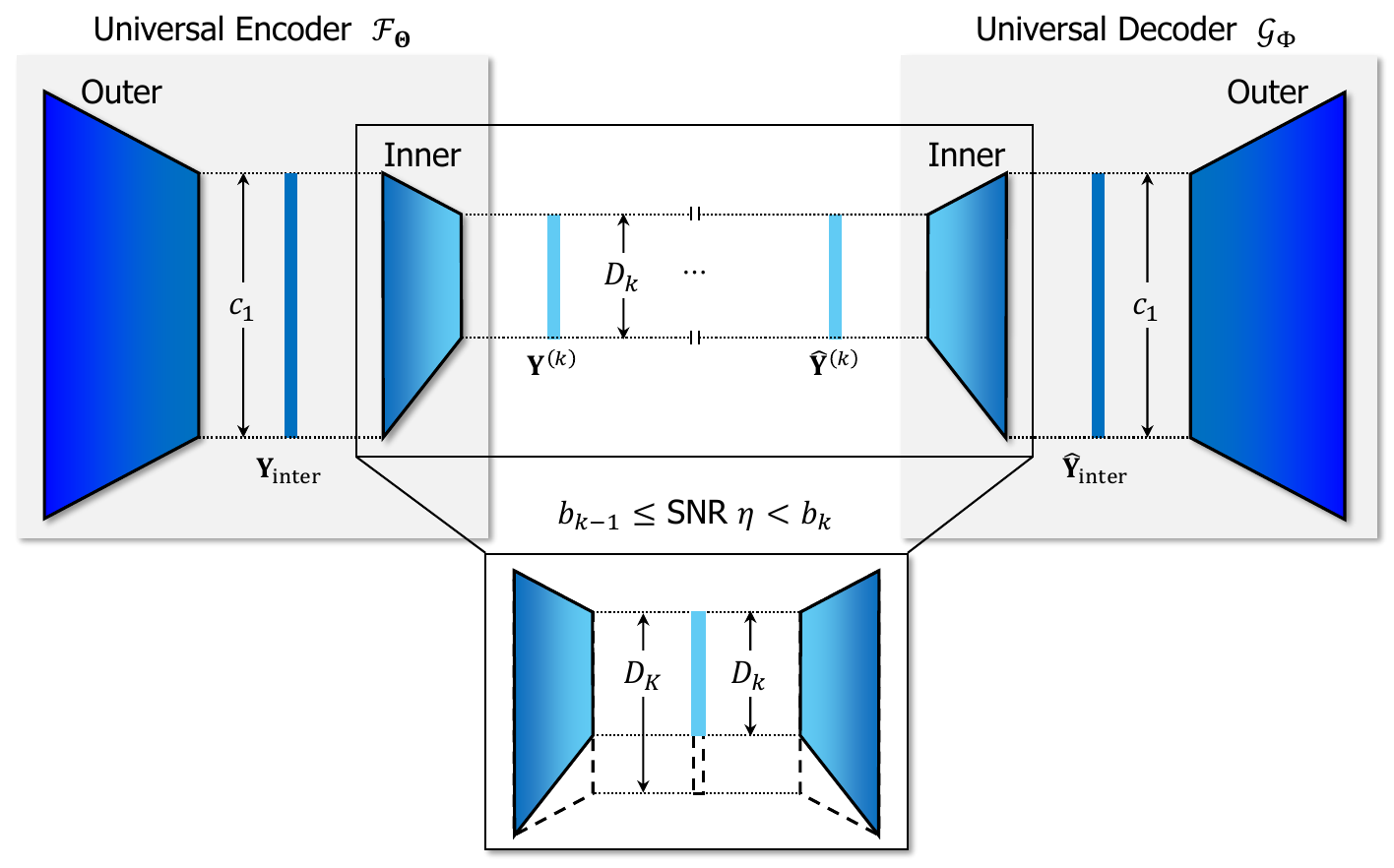}
    \vspace{-7mm}
    \caption{A detailed illustration of the conceptually divided universal encoder and decoder of uJSCC for the $k$-th modulation order, corresponding to the condition $b_{k-1} \leq \eta < b_k$.}
    \label{fig:concept}
    \vspace{-5mm}
\end{figure}

Similar to ME JSCC, the proposed modulation-agnostic semantic communication structure incorporates a uJSCC encoder $\mathcal{F}_{\mathbf{\Theta}}$, decoder $\mathcal{G}_{\mathbf{\Phi}}$, and a quantizer/dequantizer based on codebooks $\mathbf{C}^{\scriptscriptstyle(k)}$ for all $k\in[1{:}K]$. For easier understanding, the encoder and decoder can be conceptually divided into inner and outer components. Specifically, as illustrated in Fig. \ref{fig:concept}, the encoder $\mathcal{F}_{\mathbf{\Theta}}$ comprises an outer encoder $f_{\boldsymbol{\theta}_\mathrm{outer}}$ and an inner encoder $f_{\boldsymbol{\theta}_\mathrm{inner}}$, while the decoder $\mathcal{G}_{\mathbf{\Phi}}$ consists of an inner decoder $g_{\boldsymbol{\phi}_\mathrm{inner}}$ and an outer decoder $g_{\boldsymbol{\phi}_\mathrm{outer}}$. Note that the inner encoder and decoder are responsible for efficiently processing data within the shaded area at the bottom of the figure, adapting to the varying output and input shapes required by the chosen modulation order, which will be described in detail later. In contrast, the outer encoder and decoder handle the preprocessing and postprocessing of data, respectively, in support of the inner coding process. Both the outer encoder and decoder are structured with multiple CNN layers, and the inner components are each composed of a single CNN layer. The detailed NN architectures for both the encoder and decoder will be described in Section \ref{sec:Experimental Results} and features of uJSCC. Additionally, the operation of the proposed JSCC system is explained in detail below and summarized in \textbf{Algorithm \ref{alg:Universal Joint Source-Channel Coding}}.

\begin{algorithm}[!t]
\caption{Universal Joint Source-Channel Coding}\label{alg:Universal Joint Source-Channel Coding}
\small
\begin{algorithmic}[1]
    \State{\textbf{Input: }$\mathbf{X}$, $\eta$, $\mathbf{B} = [b_1, b_2, \dots, b_{K - 1}]$, $\mathcal{F}_{\mathbf{\Theta}}$, $\mathcal{G}_{\mathbf{\Phi}}$, $\mathbf{C}^{\scriptscriptstyle(k)}$ for $k\in[1{:}K]$}
    \State{Choose $m_k$ according to $\eta$ and $\mathbf{B}$ using (\ref{eq:modulation order selection})}
    \State{Encode source image $\mathbf{X}$ into feature vectors $\mathbf{Y}^{\scriptscriptstyle(k)}$ via $\mathcal{F}_{\mathbf{\Theta}}$}
    \State{Quantize $\mathbf{Y}^{\scriptscriptstyle(k)}$ into codeword indices $\mathbf{z}^{\scriptscriptstyle(k)}$ using $\mathbf{C}^{\scriptscriptstyle(k)}$}
    \State{Modulate $\mathbf{z}^{\scriptscriptstyle(k)}$ into symbol sequence $\mathbf{s}^{\scriptscriptstyle(k)}$}
    \State{Transmit $\mathbf{s}^{\scriptscriptstyle(k)}$ via AWGN channel with $\eta$}
    \State{Detect received values into $\Bar{\mathbf{s}}^{\scriptscriptstyle(k)}$ and get $\hat{\mathbf{z}}^{\scriptscriptstyle(k)}$}
    \State{Dequantize $\hat{\mathbf{z}}^{\scriptscriptstyle(k)}$ into $\hat{\mathbf{Y}}^{\scriptscriptstyle(k)}$ using $\mathbf{C}^{\scriptscriptstyle(k)}$}
    \State{Reconstruct $\hat{\mathbf{X}}^{\scriptscriptstyle(k)}$ from $\hat{\mathbf{Y}}^{\scriptscriptstyle(k)}$ via $\mathcal{G}_{\mathbf{\Phi}}$}
    \State{\textbf{Output: }$\hat{\mathbf{X}}^{\scriptscriptstyle(k)}$}
\end{algorithmic}
\end{algorithm}

The operational mechanism of the proposed uJSCC system is described in detail as follows. It is presupposed that the channel condition in SNR, denoted as $\eta$, is known in advance to both the transmitter and receiver, facilitating the predetermined selection of the modulation order $m_k$ as delineated by \eqref{eq:modulation order selection} and consequently the index $k$. Initially, at the transmitter, the source image $\mathbf{X}$ is encoded into intermediate feature vectors $\mathbf{Y}_{\mathrm{inter}} \in \mathbb{R}^{c_1 \times h_1 \times w_1}$ through the outer encoder, represented by $f_{\boldsymbol{\theta}_{\mathrm{outer}}}$, such that
\begin{align}
    \mathbf{Y}_{\mathrm{inter}} = f_{\boldsymbol{\theta}_\mathrm{outer}}(\mathbf{X}, k).
\end{align}

The intermediate vectors are subsequently processed by the inner encoder $f_{\boldsymbol{\theta}_{\mathrm{inner}}}$ tailored for the $k$-th modulation order. The inner encoder transforms these vectors into reshaped feature vectors $\mathbf{Y}^{\scriptscriptstyle(k)} = \left[\mathbf{y}_1^{\scriptscriptstyle(k)}, \mathbf{y}_2^{\scriptscriptstyle(k)}, \dots, \mathbf{y}_N^{\scriptscriptstyle(k)}\right]^T$, appropriate for $k \in [1{:}K]$, such as
\begin{align}
    \mathbf{Y}^{\scriptscriptstyle(k)} = f_{\boldsymbol{\theta}_{\mathrm{inner}}}( \mathbf{Y}_{\mathrm{inter}}, k),
\end{align}
where $\boldsymbol{\theta}_{\mathrm{inner}}$ denotes the NN parameters of the inner encoder. As illustrated in Fig. \ref{fig:concept}, the input tensor of $f_{\boldsymbol{\theta}_{\mathrm{inner}}}$ has $c_1$ channels and $f_{\boldsymbol{\theta}_{\mathrm{inner}}}$ can generate the output tensor with up to $D_K$ channels using a single CNN layer. When $b_{k - 1} \leq \eta < b_k$, corresponding to the $k$-th modulation order, the inner encoder operates with computational efficiency to yield $\mathbf{Y}^{\scriptscriptstyle(k)}$ in the desired configuration of $N \times D_k$, where  $D_k \leq D_K$, as depicted by the solid line in the figure. Such a process avoids simply slicing an output tensor of dimension $N \times D_K$, represented by the dashed line of $\mathbf{Y}^{\scriptscriptstyle(k)}$ at the bottom of the figure. Instead, it leverages the capabilities of the inner encoder to adapt the dimensions effectively for each specific modulation order.

Let $\boldsymbol{\theta}_{\mathrm{inner}}^{\scriptscriptstyle \mathrm{CNN}}\in\mathbb{R}^{D_K\times c_1\times f\times f}$ represent the tensor of CNN parameters in $\boldsymbol{\theta}_{\mathrm{inner}}$, where $f\times f$ denotes the kernel size. The tensor comprises $D_K$ convolutional filters, each of shape $c_1\times f\times f$, which are applied to a zero-padded input tensor using striding techniques. Depending on the chosen modulation order $m_k$, only the first $D_k$ filters, denoted as $\left[\boldsymbol{\theta}_{\mathrm{inner}}^{\scriptscriptstyle \mathrm{CNN}}\right]_{1:D_k, :, :, :}$, are utilized by the CNN layer of the inner encoder for encoding purposes. This corresponds to the solid-line segment of the inner encoder at the bottom of Fig. \ref{fig:concept}, while the dashed-line segment is unused. As $k$ increases, so does the dimension of the feature vector for each modulation order, with $D_1 < D_2 < \dots < D_K$, resulting in reduced computational load for lower modulation orders due to the utilization of fewer parameters. Consequently, the inner encoder requires $2N D_k c_1 f^2$ floating-point operations (FLOPs) for encoding, in contrast to $2N D_K c_1 f^2$ FLOPs when all filters are used, which is $D_K/D_k$ times more. This sophisticated encoding process achieves the same data processing efficacy as an encoder specifically designed for the $k$-th modulation order, consistent with the TE approach. The universal encoder $\mathcal{F}_{\mathbf{\Theta}}$ can be mathematically expressed as
\begin{align}\label{eq:proposed universal encoder}
    \mathcal{F}_{\mathbf{\Theta}}(\mathbf{X}, \eta) = f_{\boldsymbol{\theta}_{\mathrm{inner}}}(f_{\boldsymbol{\theta}_{\mathrm{outer}}}(\mathbf{X}, k), k).
\end{align}

Subsequently, VQ is performed using a codebook $\mathbf{C}^{\scriptscriptstyle(k)}$, which should also be optimized jointly with the encoder and decoder during training, based on the minimum distance criterion. For each $i$-th feature vector $\mathbf{y}_i^{\scriptscriptstyle(k)}$, a symbol index $z_i^{\scriptscriptstyle(k)}$ is selected according to the equation $z_i^{\scriptscriptstyle(k)} = \argmin_{j} \|\mathbf{y}_i^{\scriptscriptstyle(k)} - \mathbf{c}_j^{\scriptscriptstyle(k)}\|_2^2$. Each codeword $\mathbf{c}_j^{\scriptscriptstyle(k)}$ corresponds one-to-one to a symbol $\mathbf{s}_j^{\scriptscriptstyle(k)}$ indexed on the corresponding constellation map $\mathbb{S}^{\scriptscriptstyle(k)}$. An index vector $\mathbf{z}^{\scriptscriptstyle(k)} = \left[z_1^{\scriptscriptstyle(k)}, z_2^{\scriptscriptstyle(k)}, \dots, z_N^{\scriptscriptstyle(k)}\right]^T$ is then modulated into a symbol sequence $\mathbf{s}^{\scriptscriptstyle(k)}\in\left(\mathbb{S}^{\scriptscriptstyle(k)}\right)^N$, which is subsequently transmitted through an additive white Gaussian noise (AWGN) channel.

Upon receiving the symbol sequence altered by AWGN noise, such that $\hat{\mathbf{s}}^{\scriptscriptstyle(k)} = \mathbf{s}^{\scriptscriptstyle(k)} + \mathbf{n}$, the receiver detects it into $\Bar{\mathbf{s}}^{\scriptscriptstyle(k)}\in\left(\mathbb{S}^{\scriptscriptstyle(k)}\right)^N$. Subsequently, vector dequantization is performed using the detected index vector $\hat{\mathbf{z}}^{\scriptscriptstyle(k)} = \left[\hat{z}_1^{\scriptscriptstyle(k)}, \hat{z}_2^{\scriptscriptstyle(k)}, \dots, \hat{z}_N^{\scriptscriptstyle(k)}\right]^T$ and the same codebook $\mathbf{C}^{\scriptscriptstyle(k)}$ employed at the transmitter, yielding the received feature vectors $\hat{\mathbf{Y}}^{\scriptscriptstyle(k)}$. These vectors are then decoded by the inner decoder into recovered intermediate feature vectors
\begin{align}
    \hat{\mathbf{Y}}_{\mathrm{inter}} = g_{\boldsymbol{\phi}_{\mathrm{inner}}}(\hat{\mathbf{Y}}^{\scriptscriptstyle(k)}, k),
\end{align}
where $\boldsymbol{\phi}_{\mathrm{inner}}$ denotes the NN parameters of the inner decoder. As visualized in Fig. \ref{fig:concept}, the input tensor of $g_{\boldsymbol{\phi}_{\mathrm{inner}}}$ can have up to $D_K$ channels and $g_{\boldsymbol{\phi}_{\mathrm{inner}}}$ generate the output tensor with $c_1$ channels utilizing a single transpose CNN layer. When $b_{k - 1} \leq \eta < b_k$, corresponding to the $k$-th modulation order, the inner decoder efficiently processes $\hat{\mathbf{Y}}^{\scriptscriptstyle(k)}$ with $D_k$ ($\leq D_K$) channels, depicted by the solid line, without requiring zero-padding to conform to the shape $N\times D_K$, which is represented by the dashed line of $\hat{\mathbf{Y}}^{\scriptscriptstyle(k)}$ at the bottom.

Let $\boldsymbol{\phi}_{\mathrm{inner}}^{\scriptscriptstyle \mathrm{CNN}}\in\mathbb{R}^{D_K\times c_1\times f\times f}$ denote the tensor of CNN parameters in $\boldsymbol{\phi}_{\mathrm{inner}}$, where $f\times f$ represents the kernel size. The tensor comprises $D_K$ transpose convolution filters, each of shape $c_1\times f\times f$, while the $i$-th filter is applied to the $i$-th channel of the input tensor via the Kronecker product, followed by striding, summation, and edge cutting for size adjustment. Similarly to the encoder, the inner decoder employs only the first $D_k$ filters, denoted as $\left[\boldsymbol{\phi}_{\mathrm{inner}}^{\scriptscriptstyle \mathrm{CNN}}\right]_{1:D_k, :, :, :}$, for decoding, corresponding to the modulation order $m_k$. This is illustrated as the solid-line segment of the inner decoder at the bottom of Fig. \ref{fig:concept}, while the dashed-line segment is not utilized. The selective filter usage necessitates $2N D_k c_1 f^2$ FLOPs for lower modulation orders, in contrast to $2N D_K c_1 f^2$ FLOPs required if all filters were utilized. This optimized decoding process mirrors the efficiency of a decoder specifically tailored for the $k$-th modulation order, consistent with the TE approach.

Finally, the receiver reconstructs the image $\hat{\mathbf{X}}^{\scriptscriptstyle(k)}$ from $\hat{\mathbf{Y}}^{\mathrm{inter}}$ using the outer decoder $g_{\boldsymbol{\phi}_{\mathrm{outer}}}$, described as
\begin{align}
    \hat{\mathbf{X}}^{\scriptscriptstyle(k)} = g_{\boldsymbol{\phi}_{\mathrm{outer}}}(\hat{\mathbf{Y}}_{\mathrm{inter}}, k),
\end{align}
where $\boldsymbol{\phi}_{\mathrm{outer}}$ are the NN parameters of the outer decoder. The overall operation of the universal decoder $\mathcal{G}_{\mathbf{\Phi}}$ is encapsulated by
\begin{align}\label{eq:proposed universal decoder}
    \mathcal{G}_{\mathbf{\Phi}}(\hat{\mathbf{Y}}, \eta) = g_{\boldsymbol{\phi}_{\mathrm{outer}}}(g_{\boldsymbol{\phi}_{\mathrm{inner}}}(\hat{\mathbf{Y}}^{\scriptscriptstyle(k)}, k), k).
\end{align}

\subsection{Switchable Batch Normalization Layer}\label{subsec:Switchable Batch Normalization Layer}
Batch normalization (BN) layers \cite{ioffe2015batch} significantly facilitate the training of NN models and enhance their performance by standardizing the statistics of an output tensor from each NN layer. This ensures that subsequent layers receive inputs with consistent statistical properties during each forward pass.

Let $\mathbf{T}_{n, i}$ denote the $i$-th channel of the output tensor from the $n$-th NN layer, and let $t_{n, i, j}$ represent its $j$-th element, both pertaining to a single mini-batch. The BN layer first normalizes each $t_{n, i, j}$ into a normalized value $\hat{t}_{n, i, j}$ using the mini-batch mean $\mu_{n, i}$ and standard deviation $\sigma_{n, i}$. The normalized values are then scaled and biased to produce the batch-normalized value $\Tilde{t}_{n, i, j}$, as described by the following equations:
\begin{align}
    \hat{t}_{n, i, j} &= \frac{t_{n, i, j} - \mu_{n, i}}{\sigma_{n, i} + \epsilon},\\
    \Tilde{t}_{n, i, j} &= \gamma_{n, i}\hat{t}_{n, i, j} + \beta_{n, i},
\end{align}
where $\gamma_{n, i}$ and $\beta_{n, i}$ are the parameters learned for rescaling and bias adjustment, respectively, and $\epsilon$ is a small constant for numerical stability.

During inference, where batch-level computations are absent, the mean $\mu_{n, i}$ and standard deviation $\sigma_{n, i}$ from the training phase are replaced with running estimates $\Bar{\mu}_{n, i}$ and $\Bar{\sigma}_{n, i}$, calculated via a moving average. Simplified expressions during inference are:
\begin{align}
    \Tilde{\gamma}_{n, i} &= \frac{\gamma_{n, i}}{\Bar{\sigma}_{n, i} + \epsilon},\\
    \Tilde{\beta}_{n, i} &= \beta_{n, i} - \Tilde{\gamma}_{n, i}\Bar{\mu}_{n, i},\\
    \Tilde{t}_{n, i, j} &= \Tilde{\gamma}_{n, i}t_{n, i, j} + \Tilde{\beta}_{n, i},
\end{align}
which require only two adjusted variables per channel.

\begin{figure}[!t]
    \centering
    \includegraphics[width=\linewidth]{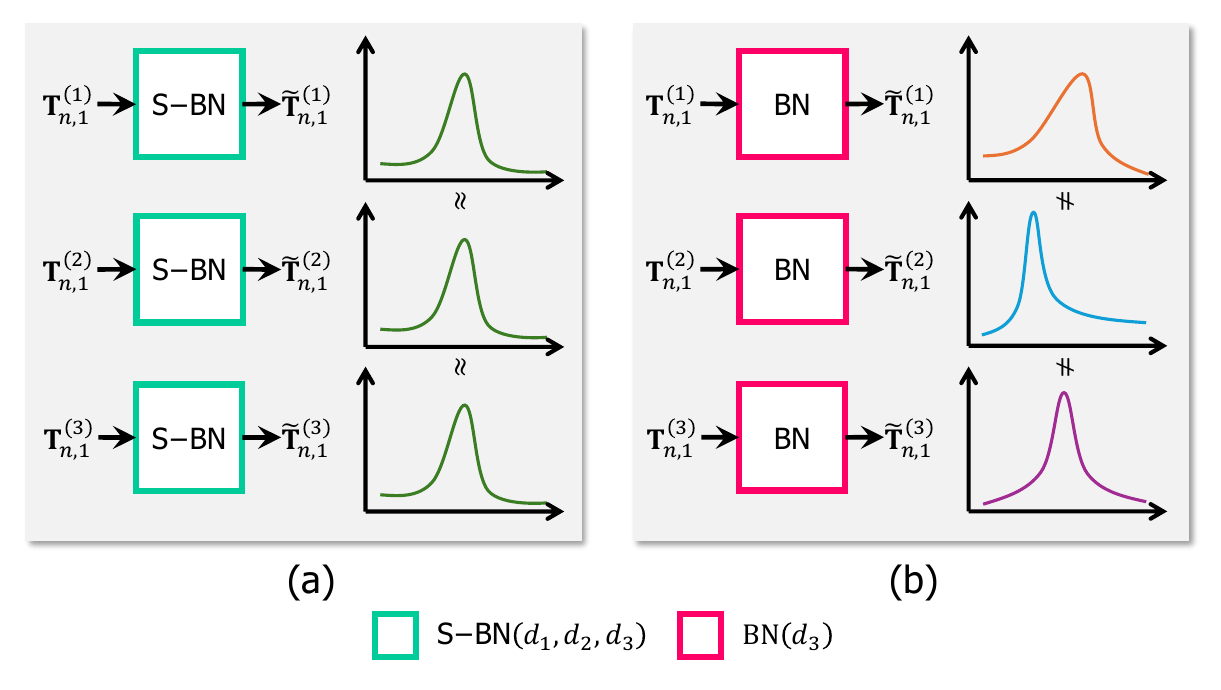}
    \vspace{-7mm}
    \caption{The processes and output probability distributions of (a) S-BN layer and (b) BN layer for the first channel of the output tensor from the $n$-th NN layer across three different paths.}
    \label{fig:statistics}
    \vspace{-5mm}
\end{figure}

The architecture of the proposed model integrates a BN layer following each CNN layer. Given that the proposed uJSCC model's input tensor size for the inner decoder $g_{\boldsymbol{\phi}_\mathrm{inner}}$ varies with each modulation order, it is essential to manage the differing statistics of the output tensor across modulation orders. To accommodate the $K$ different modulation orders in uJSCC, we employ $K$ specific BN layers, each configured to manage the statistics for its respective data processing path, thus maintaining optimal task performance despite modulation-agnostic characteristics, which is the biggest difference with the ME approach. Furthermore, the variability in statistics impacts other CNN layers, necessitating $K$ switchable BN (S-BN) layers post each CNN layer, selected based on the modulation order $m_k$. Specifically, as illustrated in Fig. \ref{fig:statistics}, where $\mathbf{T}^{\scriptscriptstyle(k)}_{n, i}$ corresponds to the tensor of the $k$-th data processing path and $d_1 < d_2 < d_3$, using the same BN layer, $\mathrm{BN}(d_3)$, results in varying output tensor statistics across the three paths. In contrast, employing the S-BN layer, $\mathrm{S{-}BN}(d_1, d_2, d_3)$ incorporating $\mathrm{BN}(d_1)$, $\mathrm{BN}(d_2)$ and $\mathrm{BN}(d_3)$, ensures greater consistency in the output tensor statistics across the paths. Moreover, this multi-layered S-BN concept, referred to in \cite{yu2018slimmable}, ensures model efficiency given that the parameter count of S-BN is minimal compared to the total model parameters.

\subsection{Training Strategy}\label{subsec:Training Strategy}
The training procedure for the uJSCC system is conducted jointly across all modulation orders. The method optimizes the universal encoder parameters $\mathbf{\Theta} = \{\boldsymbol{\theta}_{\mathrm{outer}}, \boldsymbol{\theta}_{\mathrm{inner}}\}$, universal decoder parameters $\mathbf{\Phi} = \{\boldsymbol{\phi}_{\mathrm{outer}}, \boldsymbol{\phi}_{\mathrm{inner}}\}$, and codebooks $\mathbf{C}^{\scriptscriptstyle(k)}$ for $k\in[1{:}K]$. The overall training process is outlined below and summarized in \textbf{Algorithm \ref{alg:uJSCC Training Algorithm}}.

\begin{algorithm}[!t]
\caption{uJSCC Training Algorithm}\label{alg:uJSCC Training Algorithm}
\small
\begin{algorithmic}[1]
    \State{\textbf{Input: }$\mathcal{D}$, $\mathbf{B}_\textrm{train} = [b_0, b_1, \dots, b_K]$, ($\alpha_k$, $\beta_k$, $\lambda_k$) for $k\in[1{:}K]$, $I_\textrm{max}$}
    \State{\textbf{Initialize: }$\mathbf{\Theta} = \{\boldsymbol{\theta}_\mathrm{outer}, \boldsymbol{\theta}_\mathrm{inner}\}$, $\mathbf{\Phi} = \{\boldsymbol{\phi}_\mathrm{outer}, \boldsymbol{\phi}_\mathrm{inner}\}$, $\mathbf{C}^{\scriptscriptstyle(k)}$ for $k\in[1{:}K]$}
    \State{$\boldsymbol{\varphi} \gets \{\mathbf{\Theta}, \mathbf{\Phi}, \mathbf{C}^{(1)}, \mathbf{C}^{(2)}, \dots, \mathbf{C}^{\scriptscriptstyle(K)}\}$}
    \State{$i = 0$}\Comment{\textit{Iteration index}}
    \While{$i < I_\textrm{max}$ or early stopping criterion is not met}
        \State{$\mathbf{J} = [~~]$}\Comment{\textit{Empty array}}
        \For {$k = 1{:}K$}
            \State{Sample $\eta$ such that $b_{k - 1} \leq \eta < b_k$}
            \State{Append $\eta$ into $\mathbf{J}$}
        \EndFor
        \State{$\mathcal{L}_{\text{total}} = 0$}\Comment{\textit{Clear loss value}}
        \For{$\eta$ in $\mathbf{J}$}
            \State{Choose $m_k$ by $\eta$ and $\mathbf{B}_\textrm{train}$ using (\ref{eq:modulation order selection})}
            \State{Compute $\mathcal{L}_k$ via \textbf{Algorithm \ref{alg:Universal Joint Source-Channel Coding}} and (\ref{eq:mod loss function})}
            \State{$\mathcal{L}_{\text{total}} \gets \mathcal{L}_{\text{total}} + \lambda_k\mathcal{L}_k$}
        \EndFor
        \State{Update $\boldsymbol{\varphi}$ using $\mathcal{L}_{\text{total}}$ of (\ref{eq:total loss function}) via Adam optimizer}
        \State{$i \gets i + 1$}
    \EndWhile
    \State{\textbf{Output: }$\boldsymbol{\varphi}$}
\end{algorithmic}
\end{algorithm}

The training utilizes a dataset $\mathcal{D}$, with hyperparameters such as $\alpha_k$, $\beta_k$, and $\lambda_k$, and an SNR boundary $\mathbf{B}_\textrm{train}$, which includes upper and lower bounds added to $\mathbf{B}$ for training conditions. All parameters are initialized before training begins. For each training cycle, one SNR value per modulation order for all $k\in[1{:}K]$ is sampled. The corresponding loss value $\mathcal{L}_k$ is computed for each SNR using \textbf{Algorithm \ref{alg:Universal Joint Source-Channel Coding}} according to the modulation loss function (\ref{eq:mod loss function}). The overall system is trained jointly, and the total loss is computed as
\begin{align}\label{eq:total loss function}
    \mathcal{L}_{\text{total}} = \sum_{k = 1}^K\lambda_k\mathcal{L}_k,
\end{align}
where $\lambda_k$ weighs the importance of the loss terms for each modulation order. Parameter updates are performed using the Adam optimizer \cite{kingma2014adam} based on $\mathcal{L}_{\text{total}}$. The training process continues iteratively until the maximum number of iterations is reached or an early stopping criterion is met, ensuring optimal parameter optimization across the system.

\section{Experimental Results}\label{sec:Experimental Results}
This section evaluates the uJSCC scheme's performance in an AWGN channel, assuming perfect channel knowledge at both the transmitter and receiver. The task of image reconstruction is quantified using peak SNR (PSNR) and the structural similarity index measure (SSIM) \cite{hore2010image}, with modulation schemes of BPSK, 4QAM, 16QAM, 64QAM, and 256QAM across modulation orders $K = 5$. The benchmark schemes under consideration are listed as follows:
\begin{itemize}
\item \textbf{ME$_1$}, \textbf{ME$_2$}, and \textbf{TE JSCC}: These are detailed in Section \ref{subsec:Benchmark Schemes}.
\item \textbf{DeepJSCC-Q} \cite{tung2022deepjscc}: It utilizes fixed 2-dimensional codewords based on a predefined constellation map, and its ME$_1$ variant is also included as a benchmark. Both use the same neural network structure as uJSCC for a fair comparison.
\item \textbf{SSCC}: It combines better portable graphics (BPG) \cite{bpg} for source coding with low-density parity check (LDPC) \cite{gallager1962low} for channel coding, while maintaining the same number of channel uses as uJSCC. According to the IEEE 802.11ad standard, an LDPC block length of $672$ bits is used with code rates of $1/2$, $2/3$, and $3/4$.
\end{itemize}

\begin{table*}[!t]
    \caption{NN structure for CIFAR-10 dataset (``/" is used to separately describe cases $N = 256$ and $N = 1024$.)}
    \label{tab:NN structure}
    \centering
    \begin{tabular}{|c|c|c|}
        \hline
        \textbf{Module}                         & \textbf{Layers}                                                                & \textbf{Output size}          \\ \hline
        \multirow{5}{*}{\textbf{Outer encoder}} & Conv($3, c_1, 5, 1, 2$) + S-BN($c_1, c_1, c_1, c_1, c_1$) + ReLU               & $c_1\times 32\times 32$       \\ \cline{2-3} 
                                                & Conv($c_1, c_2, 5, 1, 2$) + S-BN($c_2, c_2, c_2, c_2, c_2$) + ReLU             & $c_2\times 32\times 32$       \\ \cline{2-3} 
                                                & Res($c_2$) + ReLU                                                              & $c_2\times 32\times 32$       \\ \cline{2-3} 
                                                & Conv($c_2, c_1, 2/5, 2/1, 0/2$) + S-BN($c_1, c_1, c_1, c_1, c_1$) + ReLU       & $c_1\times 16/32\times 16/32$ \\ \cline{2-3} 
                                                & Res($c_1$) + ReLU                                                              & $c_1\times 16/32\times 16/32$ \\ \hline
        \textbf{Inner encoder}                  & Conv($c_1, D_5, 5, 1, 2$) + S-BN($D_1, D_2, D_3, D_4, D_5$) + Tanh + Reshape   & $256/1024\times D_k$          \\ \hline
        \textbf{Inner decoder}                  & Reshape + T-Conv($D_5, c_1, 5, 1, 2$) + S-BN($c_1, c_1, c_1, c_1, c_1$) + ReLU & $c_1\times 16/32\times 16/32$ \\ \hline
        \multirow{5}{*}{\textbf{Outer decoder}} & T-Res($c_1$) + ReLU                                                            & $c_1\times 16/32\times 16/32$ \\ \cline{2-3} 
                                                & T-Conv($c_1, c_2, 2/5, 2/1, 0/2$) + S-BN($c_2, c_2, c_2, c_2, c_2$) + ReLU     & $c_2\times 32\times 32$       \\ \cline{2-3} 
                                                & T-Res($c_2$) + ReLU                                                            & $c_2\times 32\times 32$       \\ \cline{2-3} 
                                                & T-Conv($c_2, c_1, 5, 1, 2$) + S-BN($c_1, c_1, c_1, c_1, c_1$) + ReLU           & $c_1\times 32\times 32$       \\ \cline{2-3} 
                                                & T-Conv($c_1, 3, 5, 1, 2$) + S-BN($3, 3, 3, 3, 3$) + Tanh                       & $3\times 32\times 32$         \\ \hline
        \multirow{2}{*}{\textbf{Res(c)}}        & Conv($c, c, 3, 1, 1$) + S-BN($c, c, c, c, c$) + ReLU                           & $c\times h\times w$           \\ \cline{2-3} 
                                                & Conv($c, c, 1, 1, 0$) + S-BN($c, c, c, c, c$)                                  & $c\times h\times w$           \\ \hline
        \multirow{2}{*}{\textbf{T-Res(c)}}      & T-Conv($c, c, 1, 1, 0$) + S-BN($c, c, c, c, c$) + ReLU                         & $c\times h\times w$           \\ \cline{2-3} 
                                                & T-Conv($c, c, 3, 1, 1$) + S-BN($c, c, c, c, c$)                                & $c\times h\times w$           \\ \hline
    \end{tabular}
    \vspace{-5mm}
\end{table*}

The NN structure for the proposed model, for sizes $N = 256$ and $N = 1024$, is detailed in TABLE \ref{tab:NN structure}. Below, the notation used within the table is clarified:
\begin{itemize}
\item $\mathrm{Conv}(c_{in}, c_{out}, f, s, p)$: Represents a 2D convolutional layer with input channel $c_{in}$, output channel $c_{out}$, kernel size $f$, stride $s$, and padding $p$.
\item $\mathrm{T{-}Conv}(c_{in}, c_{out}, f, s, p)$: A 2D transpose convolutional layer, parameterized similarly to the $\mathrm{Conv}$ layer.
\item $\mathrm{Res}(c)$ and $\mathrm{T{-}Res}(c)$: Indicate a residual layer \cite{he2016deep} and transpose residual layer, respectively, each preserving the input data's dimensions and utilizing S-BN layers.
\item $\mathrm{Reshape}$: Used for reordering dimensions and altering dimensions via merging or splitting.
\item $\mathrm{ReLU}$ and $\mathrm{Tanh}$: Activation functions using rectified linear unit and hyperbolic tangent functions, respectively.
\end{itemize}

\begin{table}[!t]
    \caption{Hyperparameters for training with CIFAR-10 dataset ($\alpha$ and $\lambda$ for each modulation are written sequentially in increasing modulation order. For ME$_2$, the first stage is for 256QAM and the second stage is for the rest of the modulation orders.)}
    \label{tab:Hyperparameters for training}
    \centering
    \begin{tabular}{|ccc|cc|}
        \hline
        \multicolumn{3}{|c|}{\textbf{Model}}                                                                                    & \multicolumn{2}{c|}{\textbf{Hyperparameters}}               \\ \hline
        \multicolumn{1}{|c|}{\textbf{Setting}}             & \multicolumn{2}{c|}{\textbf{Scheme}}                               & \multicolumn{1}{c|}{$\alpha$}            & $\lambda$        \\ \hline\hline
        \multicolumn{1}{|c|}{\multirow{5}{*}{\textbf{B}}}  & \multicolumn{2}{c|}{\textbf{uJSCC}}                                & \multicolumn{1}{c|}{$5, 4, 3, 2, 1.5$}   & $1, 1, 1, 4, 16$ \\ \cline{2-5} 
        \multicolumn{1}{|c|}{}                             & \multicolumn{2}{c|}{\textbf{ME$_1$}}                               & \multicolumn{1}{c|}{$5, 4, 3, 2, 1.5$}   & $1, 1, 1, 4, 16$ \\ \cline{2-5} 
        \multicolumn{1}{|c|}{}                             & \multicolumn{1}{c|}{\multirow{2}{*}{\textbf{ME$_2$}}} & \textbf{1} & \multicolumn{1}{c|}{$1$}                 & -                \\ \cline{3-5} 
        \multicolumn{1}{|c|}{}                             & \multicolumn{1}{c|}{}                                 & \textbf{2} & \multicolumn{1}{c|}{$5, 3, 2, 1$}        & $1, 1, 1, 1$     \\ \cline{2-5} 
        \multicolumn{1}{|c|}{}                             & \multicolumn{2}{c|}{\textbf{TE}}                                   & \multicolumn{1}{c|}{$4, 1, 1, 1, 1$}     & -                \\ \hline\hline
        \multicolumn{1}{|c|}{\multirow{5}{*}{\textbf{L}}}  & \multicolumn{2}{c|}{\textbf{uJSCC}}                                & \multicolumn{1}{c|}{$3, 2, 1, 0.7, 0.5$} & $1, 1, 2, 4, 16$ \\ \cline{2-5} 
        \multicolumn{1}{|c|}{}                             & \multicolumn{2}{c|}{\textbf{ME$_1$}}                               & \multicolumn{1}{c|}{$5, 4, 3, 2, 1.5$}   & $1, 1, 1, 4, 16$ \\ \cline{2-5} 
        \multicolumn{1}{|c|}{}                             & \multicolumn{1}{c|}{\multirow{2}{*}{\textbf{ME$_2$}}} & \textbf{1} & \multicolumn{1}{c|}{$1$}                 & -                \\ \cline{3-5} 
        \multicolumn{1}{|c|}{}                             & \multicolumn{1}{c|}{}                                 & \textbf{2} & \multicolumn{1}{c|}{$5, 3, 2, 1$}        & $1, 1, 1, 1$     \\ \cline{2-5} 
        \multicolumn{1}{|c|}{}                             & \multicolumn{2}{c|}{\textbf{TE}}                                   & \multicolumn{1}{c|}{$4, 2, 1, 1, 1$}     & -                \\ \hline\hline
        \multicolumn{1}{|c|}{\multirow{5}{*}{\textbf{MS}}} & \multicolumn{2}{c|}{\textbf{uJSCC}}                                & \multicolumn{1}{c|}{$3, 2, 1, 0.7, 0.5$} & $1, 1, 1, 4, 16$ \\ \cline{2-5} 
        \multicolumn{1}{|c|}{}                             & \multicolumn{2}{c|}{\textbf{ME$_1$}}                               & \multicolumn{1}{c|}{$5, 4, 3, 2, 1.5$}   & $1, 1, 1, 4, 16$ \\ \cline{2-5} 
        \multicolumn{1}{|c|}{}                             & \multicolumn{1}{c|}{\multirow{2}{*}{\textbf{ME$_2$}}} & \textbf{1} & \multicolumn{1}{c|}{$1$}                 & -                \\ \cline{3-5} 
        \multicolumn{1}{|c|}{}                             & \multicolumn{1}{c|}{}                                 & \textbf{2} & \multicolumn{1}{c|}{$5, 3, 2, 1$}        & $1, 1, 1, 1$     \\ \cline{2-5} 
        \multicolumn{1}{|c|}{}                             & \multicolumn{2}{c|}{\textbf{TE}}                                   & \multicolumn{1}{c|}{$1, 1, 1, 1, 1$}     & -                \\ \hline
    \end{tabular}
    \vspace{-5mm}
\end{table}

Training employs \textbf{Algorithm \ref{alg:uJSCC Training Algorithm}} with SNR boundary $\mathbf{B}_\textrm{train} = [0, 5, 12, 20, 26, 30]$. CIFAR-10 dataset\cite{krizhevsky2009learning} usage includes 50,000 training, with a 20\% validation subset, and 10,000 test images. Training parameters include an initial learning rate of 0.001 except for DeepJSCC-Q using that of 0.0008, halved every 20 epochs, a batch size of 64, and a maximum of $I_{\mathrm{max}} = 400$ epochs. Hyperparameters $\alpha$ and $\lambda$ are noted in TABLE \ref{tab:Hyperparameters for training}, with $\beta$ fixed at $0.25\alpha$.

Three simulation configurations are:
\begin{itemize}
\item \textbf{Basic} (or \textbf{B}): This setup configures the number of symbols to $N = 256$. It features codeword dimensions $(D_1, D_2, D_3, D_4, D_5) = (2, 4, 8, 12, 16)$ and NN channel counts $(c_1, c_2) = (32, 64)$.
\item \textbf{Large} (or \textbf{L}): Maintaining the same number of symbols, $N = 256$, this configuration increases the codeword dimensions to $(D_1, D_2, D_3, D_4, D_5) = (4, 8, 16, 24, 32)$ and adjusts the channel counts to $(c_1, c_2) = (64, 128)$.
\item \textbf{More Symbols} (or \textbf{MS}): This configuration sets the symbol count at $N = 1024$ and retains the codeword dimensions and channel numbers of the \textbf{Basic} setup: $(D_1, D_2, D_3, D_4, D_5) = (2, 4, 8, 12, 16)$ and $(c_1, c_2) = (32, 64)$.
\end{itemize}
Each configuration aims to evaluate the performance of the system under varying structural complexities and transmission capacities.

\begin{table}[!t]
    \caption{Number of parameters in NN models (The ratio of BN parameters is written in percentage. The ratio of the total parameters compared to uJSCC is written in multiple scales.)}
    \label{tab:Number of parameters in NN models}
    \centering
    \begin{tabular}{|cc|cc|}
        \hline
        \multicolumn{2}{|c|}{\textbf{Model}}                                 & \multicolumn{2}{c|}{\textbf{Number of parameters}}                  \\ \hline
        \multicolumn{1}{|c|}{\textbf{Setting}}             & \textbf{Scheme} & \multicolumn{1}{c|}{\textbf{BN (\%)}}   & \textbf{Total}            \\ \hline\hline
        \multicolumn{1}{|c|}{\multirow{3}{*}{\textbf{B}}}  & \textbf{uJSCC}  & \multicolumn{1}{c|}{$6514$ ($2.52$\%)}  & $258098$                  \\ \cline{2-4} 
        \multicolumn{1}{|c|}{}                             & \textbf{ME}     & \multicolumn{1}{c|}{$1318$ ($0.54$\%)}  & $252902$ ($\times 0.98$)  \\ \cline{2-4} 
        \multicolumn{1}{|c|}{}                             & \textbf{TE}     & \multicolumn{1}{c|}{$6514$ ($0.52$\%)}  & $1203634$ ($\times 4.66$) \\ \hline\hline
        \multicolumn{1}{|c|}{\multirow{3}{*}{\textbf{L}}}  & \textbf{uJSCC}  & \multicolumn{1}{c|}{$12998$ ($1.29$\%)} & $1009734$                 \\ \cline{2-4} 
        \multicolumn{1}{|c|}{}                             & \textbf{ME}     & \multicolumn{1}{c|}{$2630$ ($0.26$\%)}  & $999366$ ($\times 0.99$)  \\ \cline{2-4} 
        \multicolumn{1}{|c|}{}                             & \textbf{TE}     & \multicolumn{1}{c|}{$12998$ ($0.27$\%)} & $4753478$ ($\times 4.71$) \\ \hline\hline
        \multicolumn{1}{|c|}{\multirow{3}{*}{\textbf{MS}}} & \textbf{uJSCC}  & \multicolumn{1}{c|}{$6514$ ($1.89$\%)}  & $344114$                  \\ \cline{2-4} 
        \multicolumn{1}{|c|}{}                             & \textbf{ME}     & \multicolumn{1}{c|}{$1318$ ($0.39$\%)}  & $338918$ ($\times 0.98$)  \\ \cline{2-4} 
        \multicolumn{1}{|c|}{}                             & \textbf{TE}     & \multicolumn{1}{c|}{$6514$ ($0.40$\%)}  & $1633714$ ($\times 4.75$) \\ \hline
    \end{tabular}
    \vspace{-5mm}
\end{table}

\subsection{Model Efficiency of uJSCC}

\begin{table}[!t]
    \caption{FLOPs and training epochs of NN models (FLOPs are written in ``Encoder FLOPs/Decoder FLOPs" format and the number of epochs of TE is summed right side.)}
    \label{tab:FLOPs and training epochs of NN models}
    \centering
    \begin{tabular}{|ccc|ccc|}
        \hline
        \multicolumn{3}{|c|}{\textbf{Model}}                                                                                        & \multicolumn{3}{c|}{\textbf{Measurement}}                                                     \\ \hline
        \multicolumn{1}{|c|}{\textbf{Setting}}              & \multicolumn{1}{c|}{\textbf{Scheme}}                 & \textbf{Order} & \multicolumn{1}{c|}{\textbf{FLOPs (G)}} & \multicolumn{2}{c|}{\textbf{Epochs}}                \\ \hline\hline
        \multicolumn{1}{|c|}{\multirow{10}{*}{\textbf{B}}}  & \multicolumn{1}{c|}{\multirow{5}{*}{\textbf{uJSCC}}} & BPSK           & \multicolumn{1}{c|}{$0.205/0.205$}      & \multicolumn{2}{c|}{\multirow{5}{*}{$178$}}         \\ \cline{3-4}
        \multicolumn{1}{|c|}{}                              & \multicolumn{1}{c|}{}                                & 4QAM           & \multicolumn{1}{c|}{$0.205/0.206$}      & \multicolumn{2}{c|}{}                               \\ \cline{3-4}
        \multicolumn{1}{|c|}{}                              & \multicolumn{1}{c|}{}                                & 16QAM          & \multicolumn{1}{c|}{$0.207/0.207$}      & \multicolumn{2}{c|}{}                               \\ \cline{3-4}
        \multicolumn{1}{|c|}{}                              & \multicolumn{1}{c|}{}                                & 64QAM          & \multicolumn{1}{c|}{$0.209/0.209$}      & \multicolumn{2}{c|}{}                               \\ \cline{3-4}
        \multicolumn{1}{|c|}{}                              & \multicolumn{1}{c|}{}                                & 256QAM         & \multicolumn{1}{c|}{$0.210/0.210$}      & \multicolumn{2}{c|}{}                               \\ \cline{2-6} 
        \multicolumn{1}{|c|}{}                              & \multicolumn{1}{c|}{\multirow{5}{*}{\textbf{TE}}}    & BPSK           & \multicolumn{1}{c|}{$0.205/0.205$}      & \multicolumn{1}{c|}{$41$}  & \multirow{5}{*}{$473$} \\ \cline{3-5}
        \multicolumn{1}{|c|}{}                              & \multicolumn{1}{c|}{}                                & 4QAM           & \multicolumn{1}{c|}{$0.205/0.206$}      & \multicolumn{1}{c|}{$44$}  &                        \\ \cline{3-5}
        \multicolumn{1}{|c|}{}                              & \multicolumn{1}{c|}{}                                & 16QAM          & \multicolumn{1}{c|}{$0.207/0.207$}      & \multicolumn{1}{c|}{$93$}  &                        \\ \cline{3-5}
        \multicolumn{1}{|c|}{}                              & \multicolumn{1}{c|}{}                                & 64QAM          & \multicolumn{1}{c|}{$0.209/0.209$}      & \multicolumn{1}{c|}{$143$} &                        \\ \cline{3-5}
        \multicolumn{1}{|c|}{}                              & \multicolumn{1}{c|}{}                                & 256QAM         & \multicolumn{1}{c|}{$0.210/0.210$}      & \multicolumn{1}{c|}{$152$} &                        \\ \hline\hline
        \multicolumn{1}{|c|}{\multirow{10}{*}{\textbf{L}}}  & \multicolumn{1}{c|}{\multirow{5}{*}{\textbf{uJSCC}}} & BPSK           & \multicolumn{1}{c|}{$0.807/0.807$}      & \multicolumn{2}{c|}{\multirow{5}{*}{178}}           \\ \cline{3-4}
        \multicolumn{1}{|c|}{}                              & \multicolumn{1}{c|}{}                                & 4QAM           & \multicolumn{1}{c|}{$0.811/0.811$}      & \multicolumn{2}{c|}{}                               \\ \cline{3-4}
        \multicolumn{1}{|c|}{}                              & \multicolumn{1}{c|}{}                                & 16QAM          & \multicolumn{1}{c|}{$0.817/0.817$}      & \multicolumn{2}{c|}{}                               \\ \cline{3-4}
        \multicolumn{1}{|c|}{}                              & \multicolumn{1}{c|}{}                                & 64QAM          & \multicolumn{1}{c|}{$0.824/0.824$}      & \multicolumn{2}{c|}{}                               \\ \cline{3-4}
        \multicolumn{1}{|c|}{}                              & \multicolumn{1}{c|}{}                                & 256QAM         & \multicolumn{1}{c|}{$0.830/0.830$}      & \multicolumn{2}{c|}{}                               \\ \cline{2-6} 
        \multicolumn{1}{|c|}{}                              & \multicolumn{1}{c|}{\multirow{5}{*}{\textbf{TE}}}    & BPSK           & \multicolumn{1}{c|}{$0.807/0.807$}      & \multicolumn{1}{c|}{$34$}  & \multirow{5}{*}{$441$} \\ \cline{3-5}
        \multicolumn{1}{|c|}{}                              & \multicolumn{1}{c|}{}                                & 4QAM           & \multicolumn{1}{c|}{$0.811/0.811$}      & \multicolumn{1}{c|}{$32$}  &                        \\ \cline{3-5}
        \multicolumn{1}{|c|}{}                              & \multicolumn{1}{c|}{}                                & 16QAM          & \multicolumn{1}{c|}{$0.817/0.817$}      & \multicolumn{1}{c|}{$65$}  &                        \\ \cline{3-5}
        \multicolumn{1}{|c|}{}                              & \multicolumn{1}{c|}{}                                & 64QAM          & \multicolumn{1}{c|}{$0.824/0.824$}      & \multicolumn{1}{c|}{$125$} &                        \\ \cline{3-5}
        \multicolumn{1}{|c|}{}                              & \multicolumn{1}{c|}{}                                & 256QAM         & \multicolumn{1}{c|}{$0.830/0.830$}      & \multicolumn{1}{c|}{$185$} &                        \\ \hline\hline
        \multicolumn{1}{|c|}{\multirow{10}{*}{\textbf{MS}}} & \multicolumn{1}{c|}{\multirow{5}{*}{\textbf{uJSCC}}} & BPSK           & \multicolumn{1}{c|}{$0.324/0.324$}      & \multicolumn{2}{c|}{\multirow{5}{*}{$141$}}         \\ \cline{3-4}
        \multicolumn{1}{|c|}{}                              & \multicolumn{1}{c|}{}                                & 4QAM           & \multicolumn{1}{c|}{$0.327/0.327$}      & \multicolumn{2}{c|}{}                               \\ \cline{3-4}
        \multicolumn{1}{|c|}{}                              & \multicolumn{1}{c|}{}                                & 16QAM          & \multicolumn{1}{c|}{$0.334/0.334$}      & \multicolumn{2}{c|}{}                               \\ \cline{3-4}
        \multicolumn{1}{|c|}{}                              & \multicolumn{1}{c|}{}                                & 64QAM          & \multicolumn{1}{c|}{$0.340/0.340$}      & \multicolumn{2}{c|}{}                               \\ \cline{3-4}
        \multicolumn{1}{|c|}{}                              & \multicolumn{1}{c|}{}                                & 256QAM         & \multicolumn{1}{c|}{$0.347/0.347$}      & \multicolumn{2}{c|}{}                               \\ \cline{2-6} 
        \multicolumn{1}{|c|}{}                              & \multicolumn{1}{c|}{\multirow{5}{*}{\textbf{TE}}}    & BPSK           & \multicolumn{1}{c|}{$0.324/0.324$}      & \multicolumn{1}{c|}{$64$}  & \multirow{5}{*}{$513$} \\ \cline{3-5}
        \multicolumn{1}{|c|}{}                              & \multicolumn{1}{c|}{}                                & 4QAM           & \multicolumn{1}{c|}{$0.327/0.327$}      & \multicolumn{1}{c|}{$93$}  &                        \\ \cline{3-5}
        \multicolumn{1}{|c|}{}                              & \multicolumn{1}{c|}{}                                & 16QAM          & \multicolumn{1}{c|}{$0.334/0.334$}      & \multicolumn{1}{c|}{$100$} &                        \\ \cline{3-5}
        \multicolumn{1}{|c|}{}                              & \multicolumn{1}{c|}{}                                & 64QAM          & \multicolumn{1}{c|}{$0.340/0.340$}      & \multicolumn{1}{c|}{$149$} &                        \\ \cline{3-5}
        \multicolumn{1}{|c|}{}                              & \multicolumn{1}{c|}{}                                & 256QAM         & \multicolumn{1}{c|}{$0.347/0.347$}      & \multicolumn{1}{c|}{$107$} &                        \\ \hline
    \end{tabular}
    \vspace{-5mm}
\end{table}

TABLE \ref{tab:Number of parameters in NN models} shows that BN layer parameters account for only 1-2\% of total model parameters, indicating that S-BN layers in uJSCC minimally impact overall model size, with ME$_1$ and ME$_2$ comprising about 98-99\% of uJSCC's parameters. Conversely, TE's parameter count is nearly fivefold higher due to its five separate JSCC encoder-decoder pairs for each modulation order, underscoring the proposed model's parameter efficiency.

TABLE \ref{tab:FLOPs and training epochs of NN models} details the FLOPs for the encoder and decoder during forward pass, primarily affected by the CNN layers, with nearly identical figures for uJSCC's universal encoder and decoder. FLOPs for uJSCC are on par with TE, indicating comparable operational efficiency across all of the modulation orders. uJSCC, employing parameter sharing and joint training across modulation orders, requires far fewer training epochs than TE, which trains each modulation order separately. This efficient training approach of uJSCC leads to fewer epochs especially as TE's training epochs increase with higher modulation orders due to greater codebook cardinality and longer codeword dimensions, highlighting uJSCC’s efficiency in training complexity.

\subsection{uJSCC vs. Benchmarks under \textbf{Basic} Setting}

\begin{figure}[!t]
    \centering
    \subfigure[SSIM vs. SNR]{\includegraphics[width=0.85\linewidth]{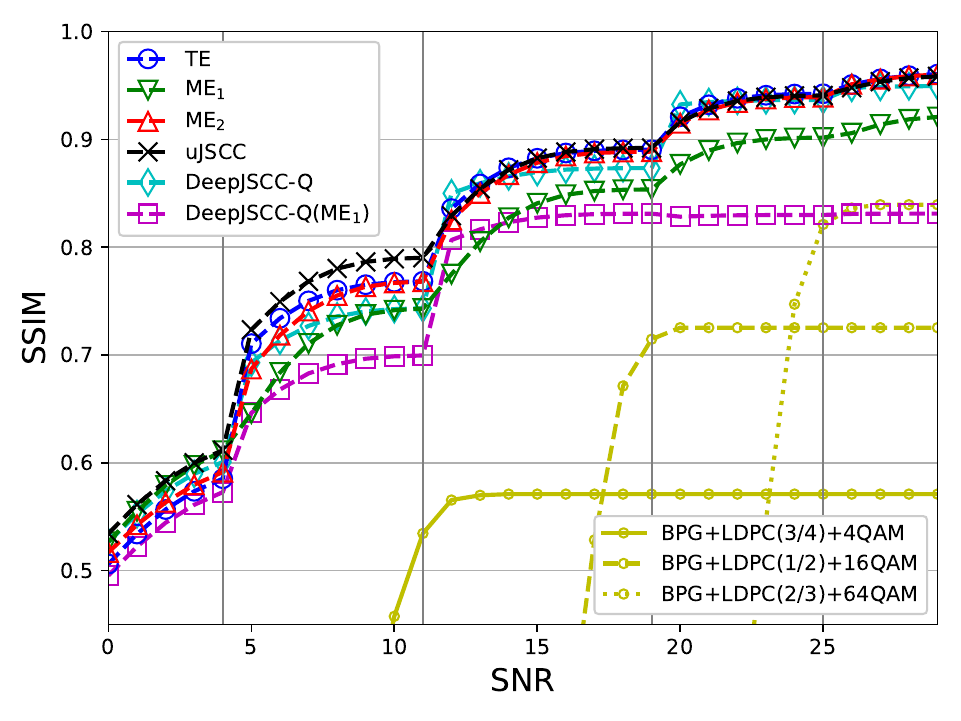}}
    \subfigure[PSNR vs. SNR]{\includegraphics[width=0.85\linewidth]{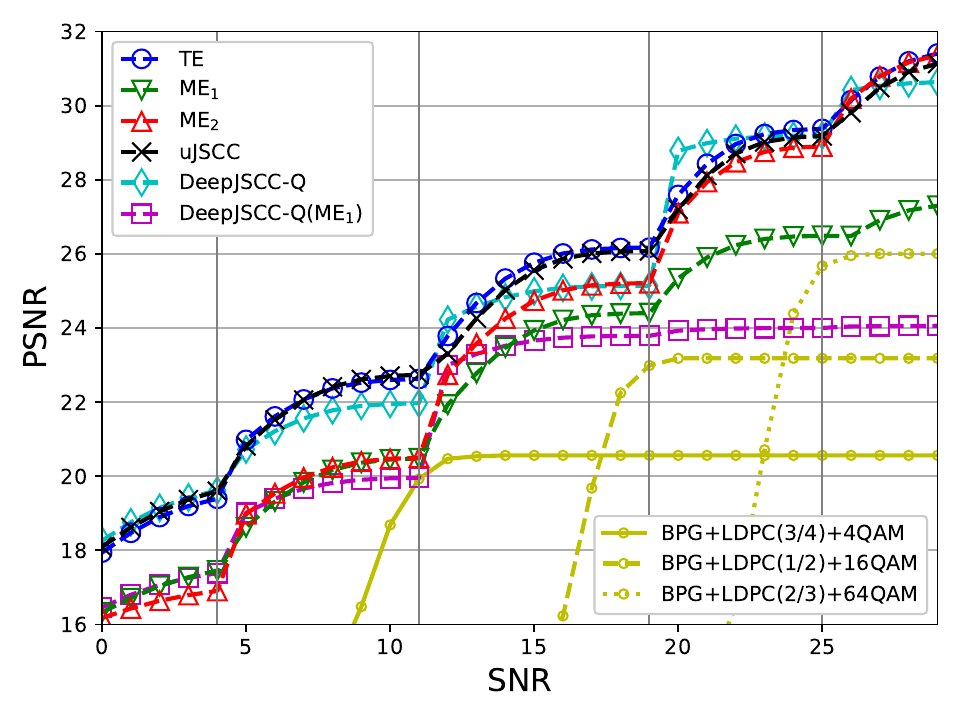}}
    \caption{Performance comparison between uJSCC vs. benchmark schemes under \textbf{Basic} setting.}
    \label{fig:Performance comparison with benchmark schemes N=256}
    \vspace{-5mm}
\end{figure}

\begin{figure}[!t]
    \centering
    \subfigure[Original image]{\includegraphics[width=0.8\linewidth]{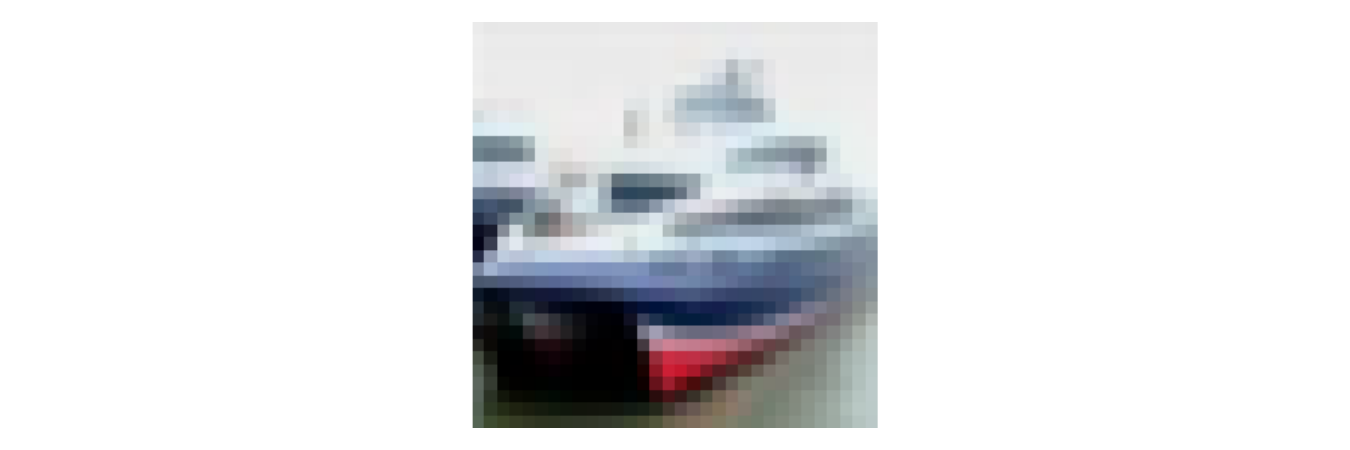}}
    \subfigure[uJSCC]{\includegraphics[width=0.9\linewidth]{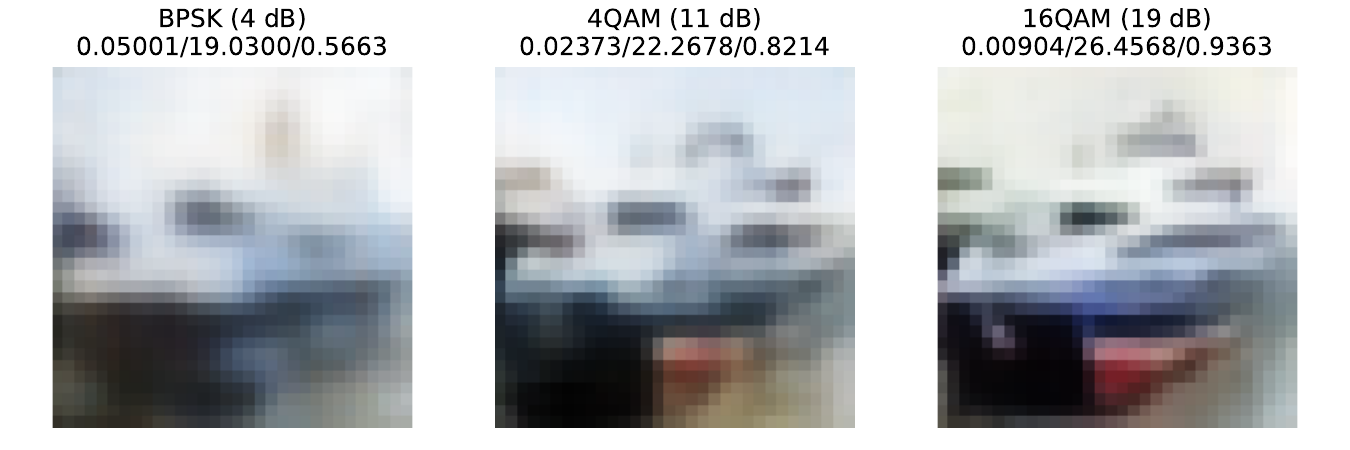}}
    \subfigure[DeepJSCC-Q]{\includegraphics[width=0.9\linewidth]{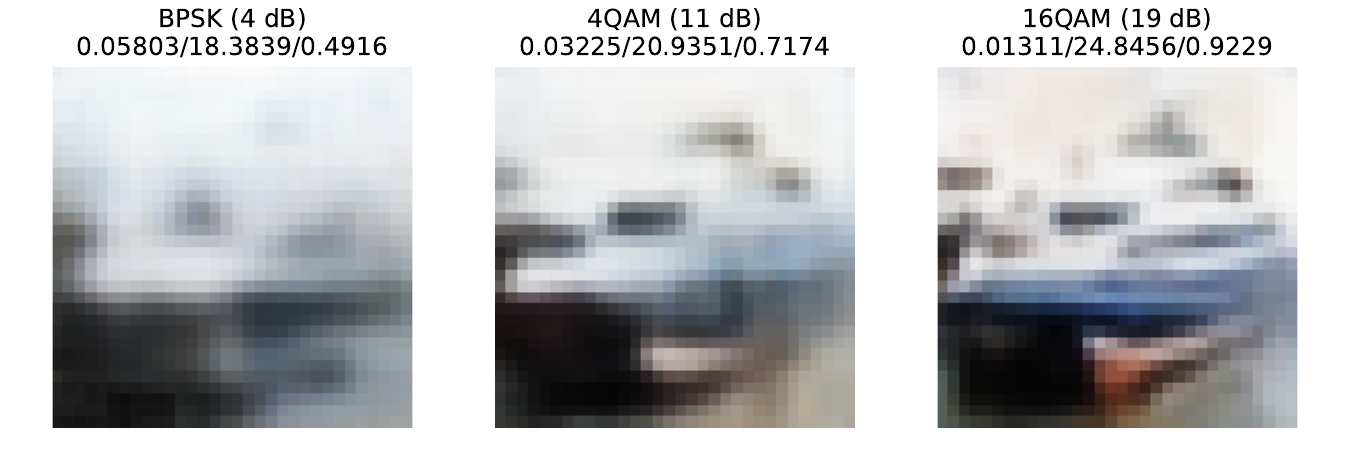}}
    \subfigure[TE]{\includegraphics[width=0.9\linewidth]{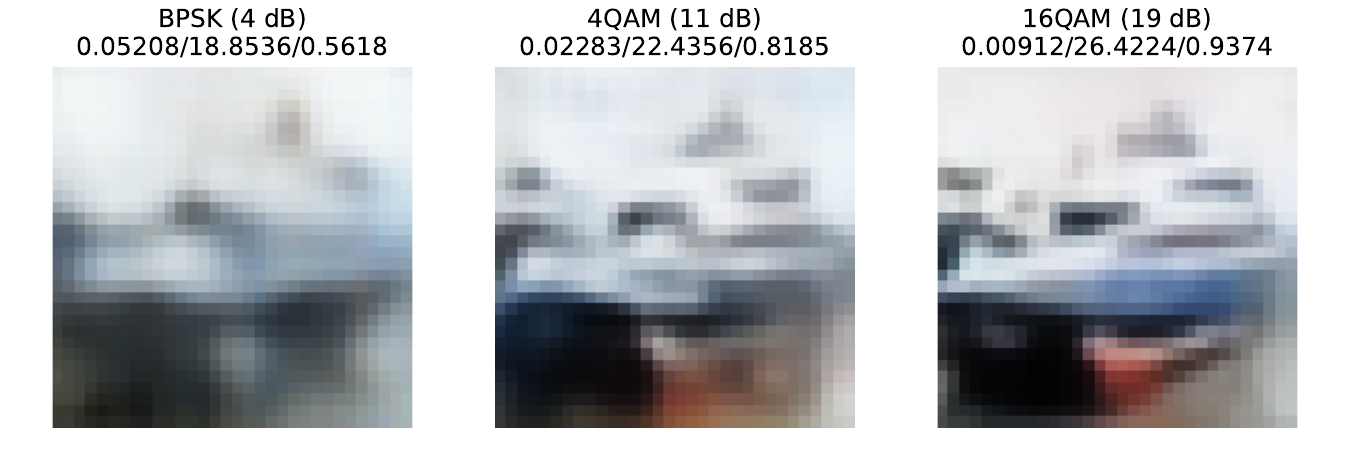}}
    \subfigure[ME$_1$]{\includegraphics[width=0.9\linewidth]{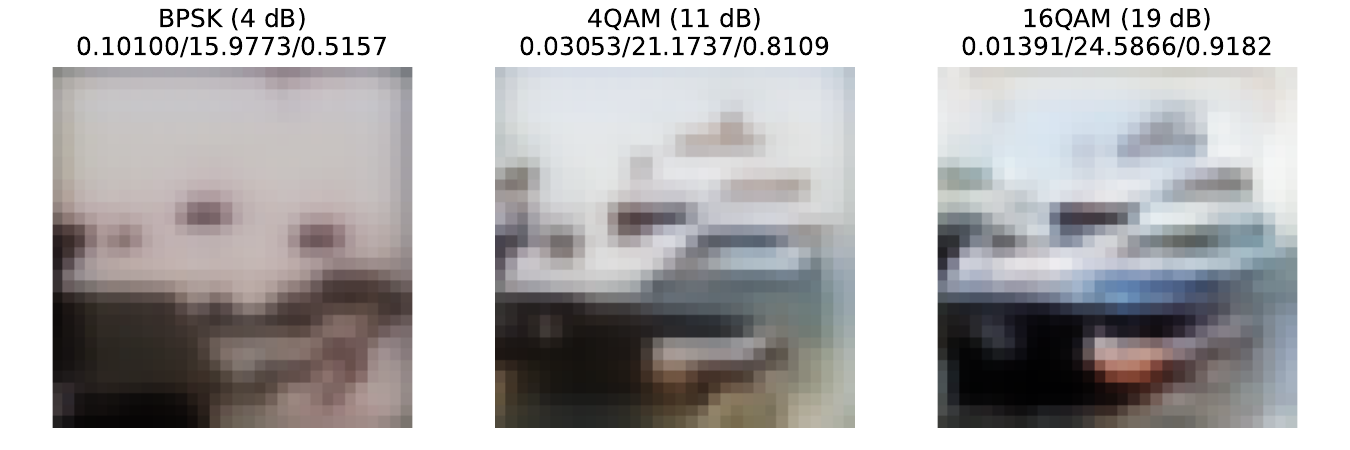}}
    \subfigure[ME$_2$]{\includegraphics[width=0.9\linewidth]{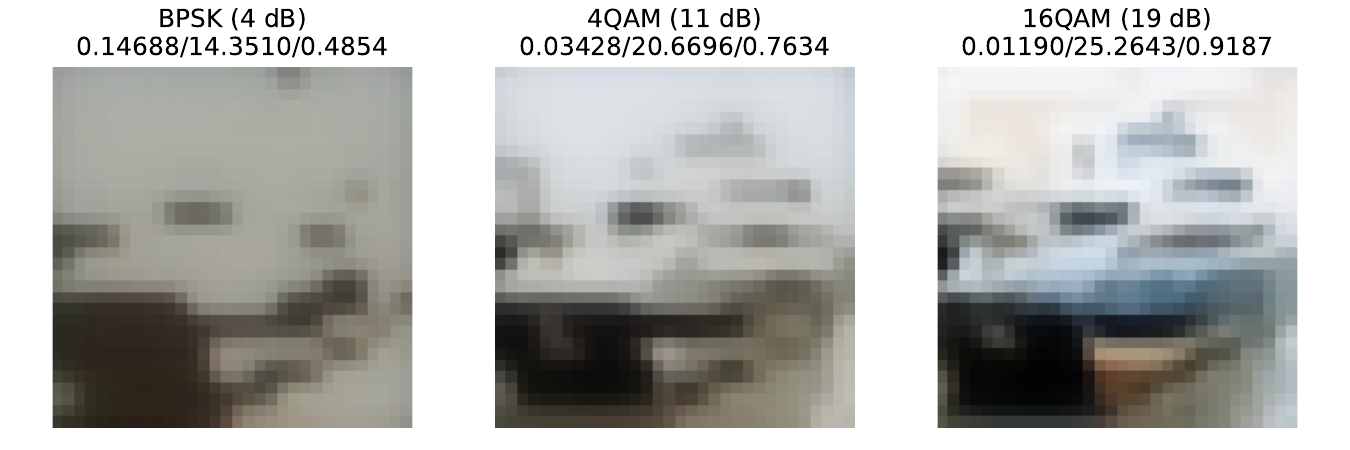}}
    \caption{Original image and reconstructed images for BPSK, 4QAM, and 16QAM under \textbf{Basic} setting. (Performance is numerically written in ``MSE/PSNR/SSIM" format.)}
    \label{fig:Reconstructed images for each modulation order N=256}
    \vspace{-5mm}
\end{figure}

Fig. \ref{fig:Performance comparison with benchmark schemes N=256} demonstrates that uJSCC, with only a single universal encoder and decoder, surpasses benchmark schemes in low SNR environments when transmitting 256 symbols using BPSK and 4QAM. It highlights the advantage of semantic communication where uJSCC, benefiting from joint training and parameter sharing across multiple modulation orders, enhances image reconstruction in lower modulation orders using the capabilities of higher orders. Notably, the graph exhibits a wave-like pattern, rather than a smooth curve, because the performance converges at high SNR values within the range of each modulation order. Therefore, as the performance degrades due to channel distortion, the graph displays a concatenated form of five distinct curves. ME$_1$ lacks S-BN layers, impairing its ability to equalize output tensor statistics across CNN layers for each modulation order. For higher orders like 64QAM and 256QAM, TE and ME$_2$ excel, with ME$_2$ matching TE's performance initially trained for 256QAM, yet showing weaker performance in lower orders due to suboptimal training across remaining codebooks. The performance of DeepJSCC-Q is evaluated using a model trained for each modulation order within its corresponding SNR range, similar to TE. DeepJSCC-Q outperforms TE at low SNR values in the range of 64QAM and 256QAM. However, as its variant in ME$_1$ style shows poorer performance than uJSCC, DeepJSCC-Q is not appropriate for a single model for supporting multiple modulation orders.

Fig. \ref{fig:Reconstructed images for each modulation order N=256} displays reconstructed images from various models for BPSK, 4QAM, and 16QAM. The results for 64QAM and 256QAM are omitted as all models, except ME$_1$, demonstrate satisfactory image quality. uJSCC clearly outperforms other models in lower modulation orders like BPSK and 4QAM. For 16QAM, uJSCC and TE show comparable and superior image quality over DeepJSCC-Q, ME$_1$ and ME$_2$, affirming uJSCC's advantage in low SNR settings through effective semantic communication.

\subsection{uJSCC vs. Benchmarks under \textbf{Large} Setting}

\begin{figure}[!t]
    \centering
    \subfigure[SSIM vs. SNR]{\includegraphics[width=0.85\linewidth]{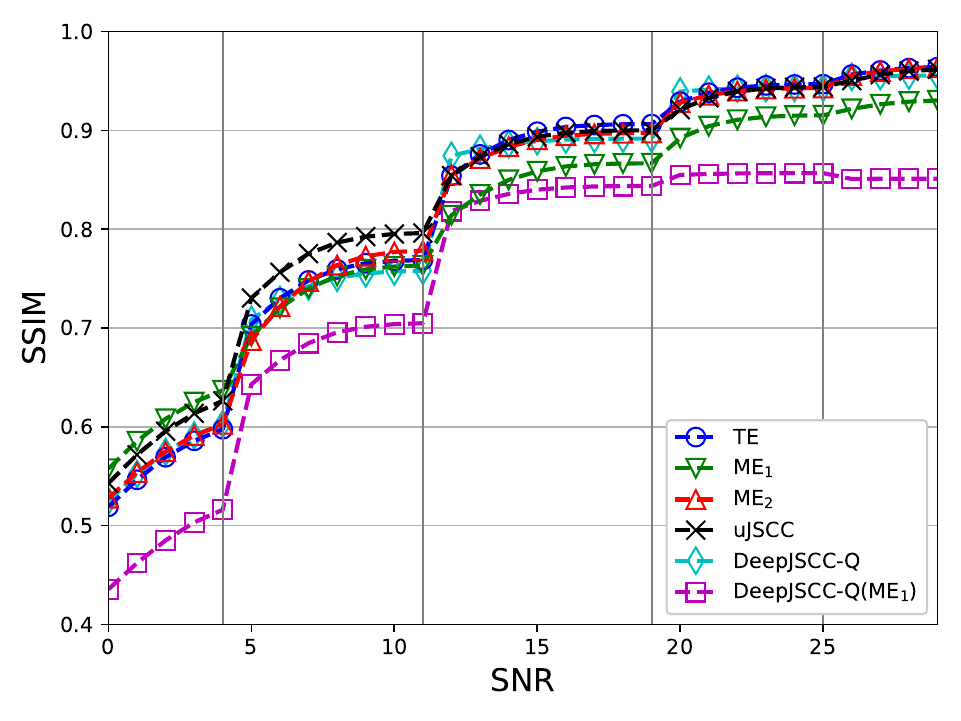}}
    \subfigure[PSNR vs. SNR]{\includegraphics[width=0.85\linewidth]{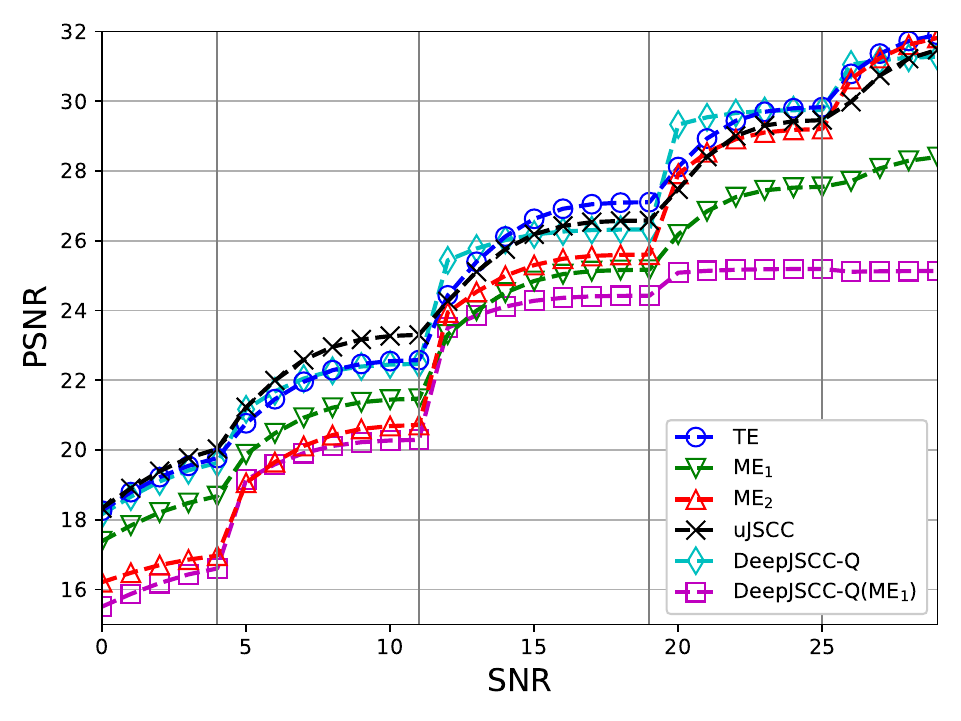}}
    \caption{Performance comparison between uJSCC vs. benchmark schemes under \textbf{Large} setting.}
    \label{fig:Performance comparison with benchmark schemes bigger size model}
    \vspace{-5mm}
\end{figure}

Fig. \ref{fig:Performance comparison with benchmark schemes bigger size model} depicts the image reconstruction performance of uJSCC and benchmark schemes using an enlarged model. This setup exhibits a trend similar to the \textbf{Basic} configuration. According to the SSIM metric, uJSCC generally outperforms except in the BPSK range, where ME$_1$ leads narrowly but shows weaker performance across other modulation orders. Notably, uJSCC significantly surpasses other models in the 4QAM range, crucial for the visual quality of the reconstructed image. 

In terms of the PSNR metric, uJSCC excels in the low SNR zones like BPSK and 4QAM and closely matches TE's performance in higher modulation orders, despite TE having approximately five times more parameters. The slight PSNR shortfall in the high SNR region is not detrimental as uJSCC nearly reaches peak performance in terms of SSIM, which is more reflective of the structural integrity of the reconstructed image. Given that ME$_1$ underperforms compared to uJSCC in the BPSK range, both SSIM and PSNR metrics are crucial for a comprehensive evaluation of the image reconstruction quality. This simulation confirms that a larger uJSCC model, when implemented with the devised training strategy, can achieve superior or competitive reconstruction across all SNR ranges with reduced training complexity relative to TE.

\subsection{uJSCC vs. Benchmarks under \textbf{More Symbols} Setting}

\begin{figure}[!t]
    \centering
    \subfigure[SSIM vs. SNR]{\includegraphics[width=0.85\linewidth]{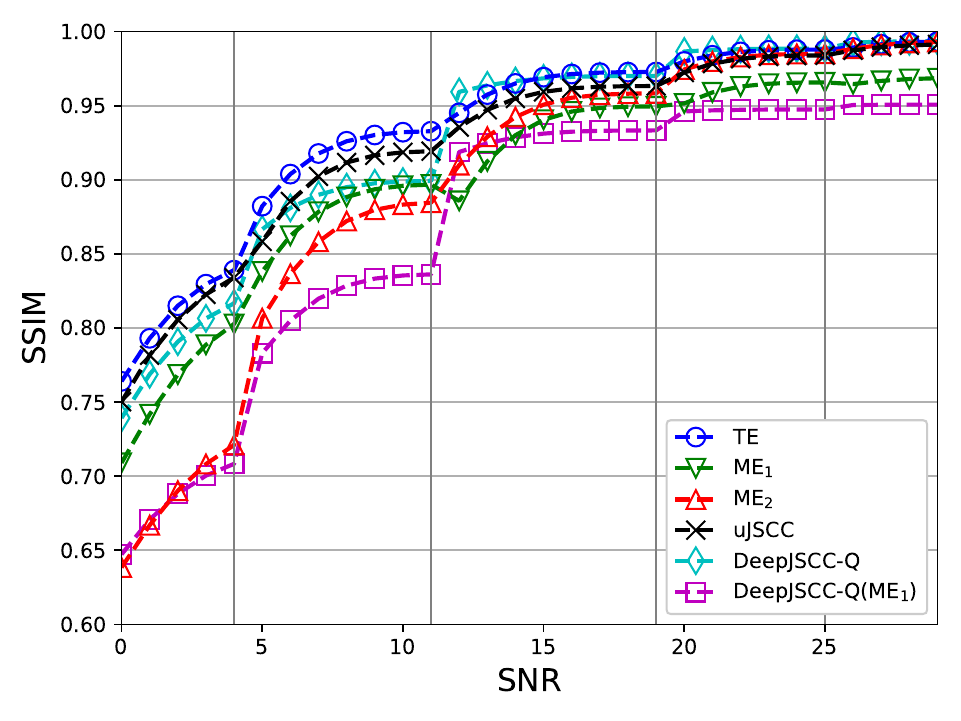}}
    \subfigure[PSNR vs. SNR]{\includegraphics[width=0.85\linewidth]{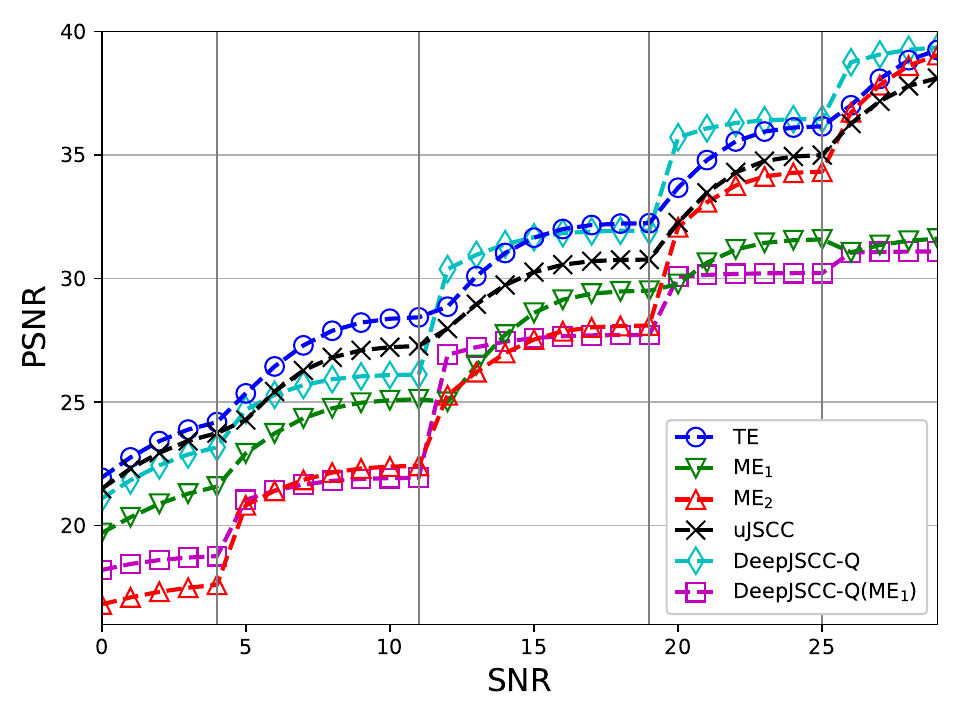}}
    \caption{Performance comparison between uJSCC vs. benchmark schemes under \textbf{More Symbols} setting.}
    \label{fig:Performance comparison with benchmark schemes N=1024}
    \vspace{-5mm}
\end{figure}

\begin{figure}[!t]
    \centering
    \subfigure[Original image]{\includegraphics[width=0.8\linewidth]{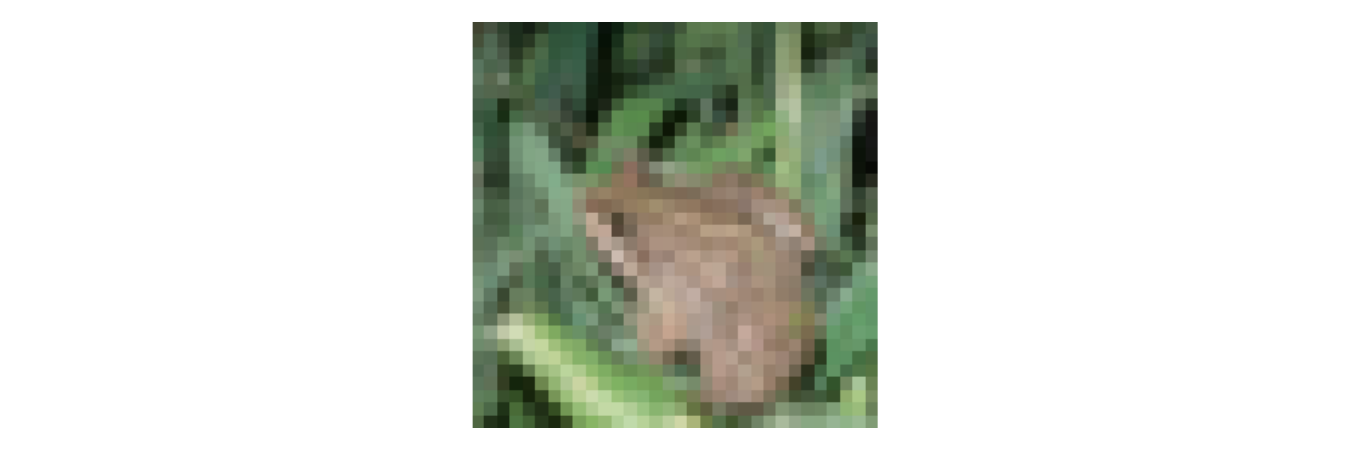}}
    \subfigure[uJSCC]{\includegraphics[width=0.9\linewidth]{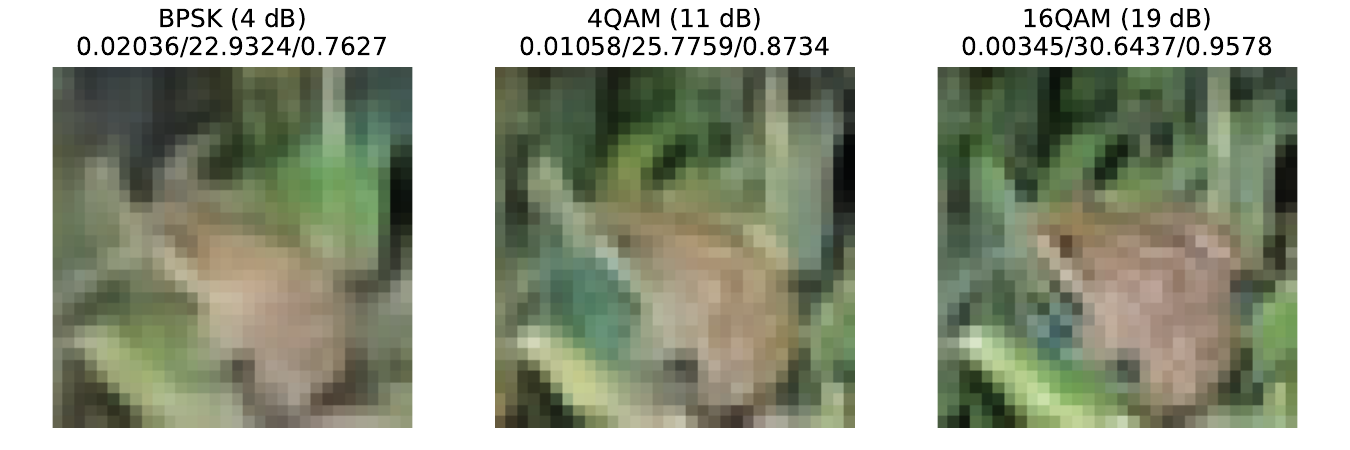}}
    \subfigure[DeepJSCC-Q]{\includegraphics[width=0.9\linewidth]{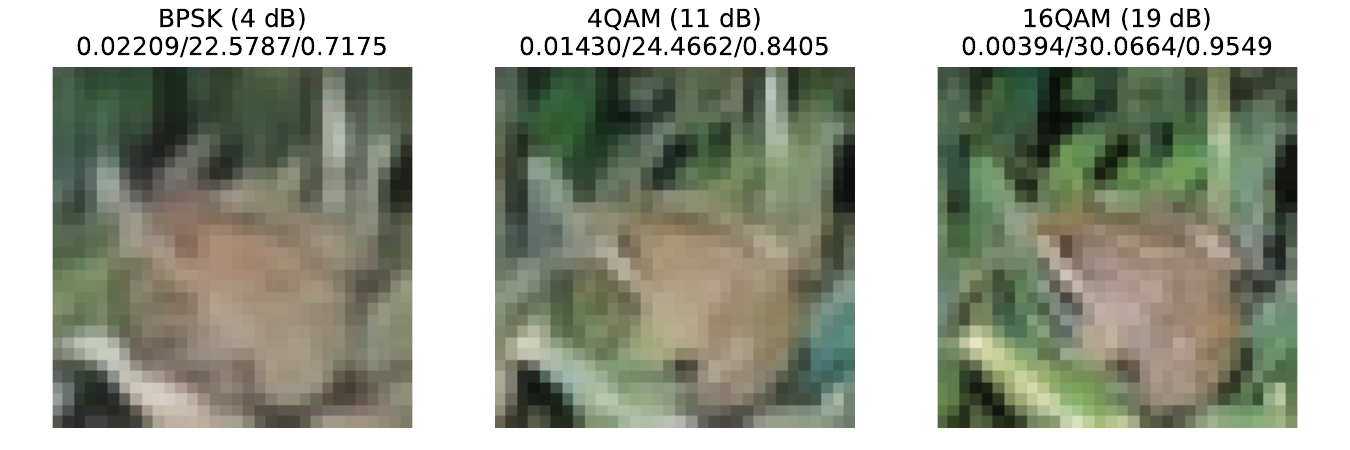}}
    \subfigure[TE]{\includegraphics[width=0.9\linewidth]{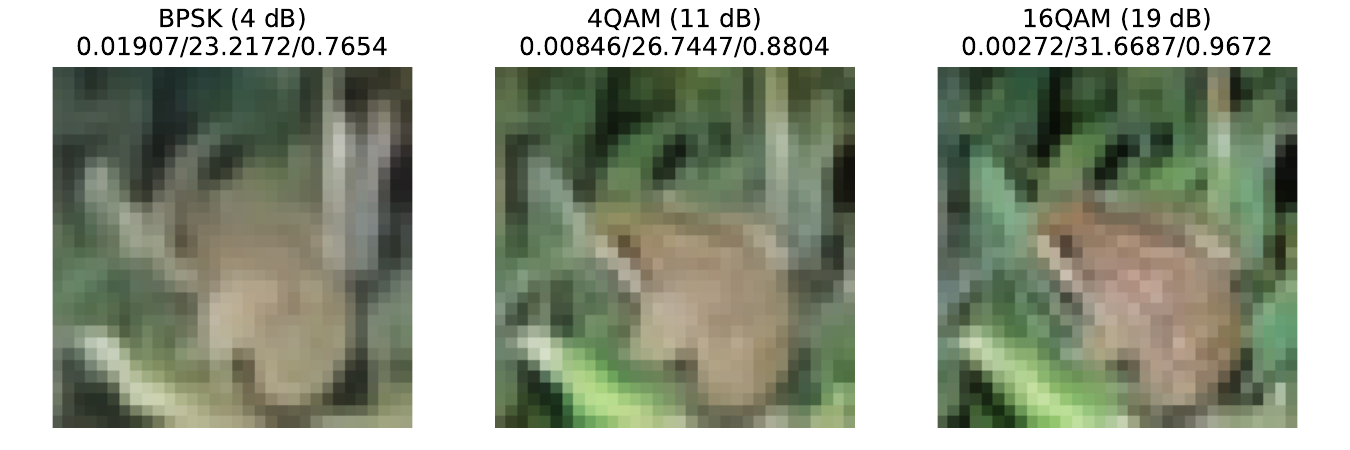}}
    \subfigure[ME$_1$]{\includegraphics[width=0.9\linewidth]{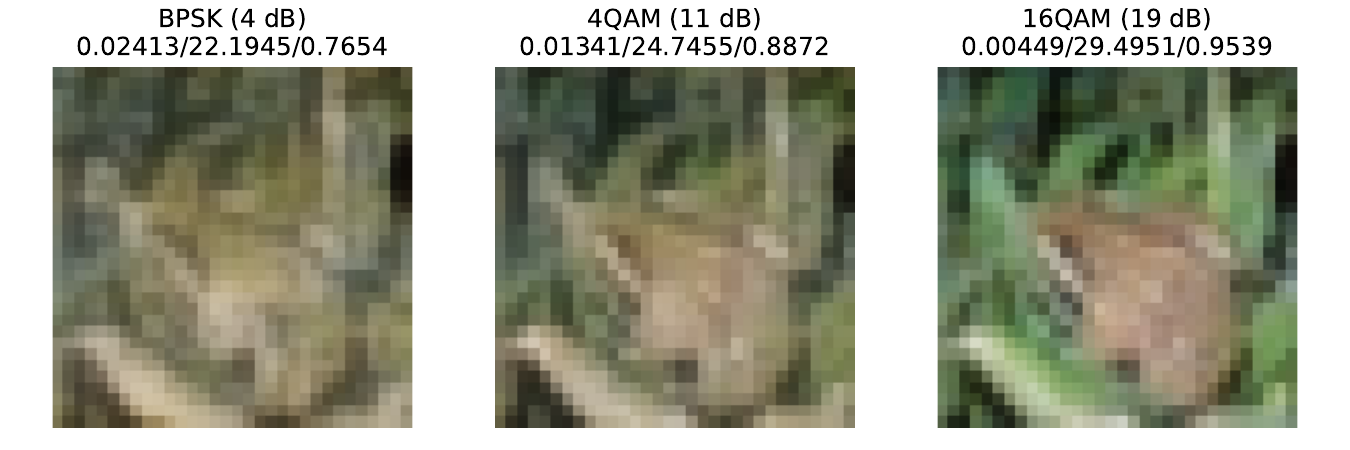}}
    \subfigure[ME$_2$]{\includegraphics[width=0.9\linewidth]{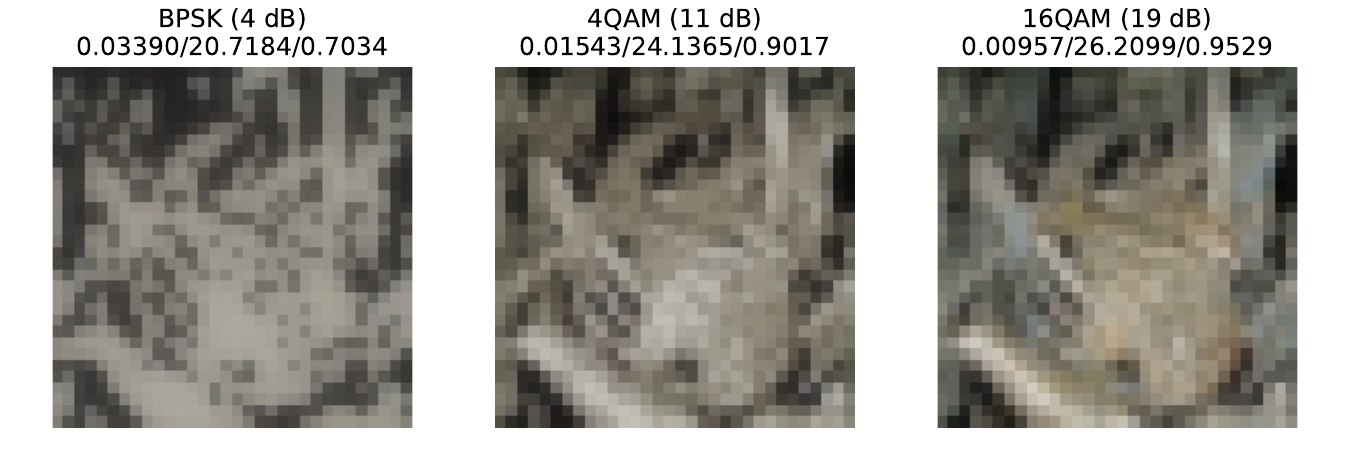}}
    \caption{Original image and reconstructed images for BPSK, 4QAM, and 16QAM under \textbf{More Symbols} setting. (Performance is numerically written in ``MSE/PSNR/SSIM" format.)}
    \label{fig:Reconstructed images for each modulation order N=1024}
    \vspace{-5mm}
\end{figure}

Fig. \ref{fig:Performance comparison with benchmark schemes N=1024} evaluates the image reconstruction performance of uJSCC and benchmark schemes with a transmission of 1024 symbols, i.e., $N = 1024$. Transmitting more symbols enhances overall performance compared to the \textbf{Basic} setting. Likewise in the \textbf{Basic} setting, although DeepJSCC-Q outperforms TE for 64QAM and 256QAM, it is not suitable as a universal model when considering its ME$_1$ version. TE excels across all SNR ranges than uJSCC, as its multiple encoder-decoder pairs are specifically trained for individual modulation orders at corresponding SNR levels. However, uJSCC closely rivals TE, particularly matching its performance in the high SNR region with 64QAM and 256QAM, as measured by SSIM. ME$_1$ consistently underperforms across all SNR ranges, while ME$_2$ performs poorly in low SNR settings with BPSK and 4QAM but well in higher SNR areas. This discrepancy highlights the challenge of optimizing a model across all SNR ranges through joint training, which is not the case for uJSCC. The effectiveness of the proposed NN structure, particularly with increased codeword dimension order and the integration of S-BN layers, alongside the training algorithm, is demonstrated in achieving high-quality outcomes for the downstream task. uJSCC maintains moderate to high performance across a broad SNR spectrum, supporting multiple modulation orders with a single universal model.

Fig. \ref{fig:Reconstructed images for each modulation order N=1024} presents the reconstructed images from each model for BPSK, 4QAM, and 16QAM when transmitting 1024 symbols. As higher modulation orders like 64QAM and 256QAM generally yield satisfactory results, the focus is on lower orders. ME$_2$, optimized only for 256QAM, shows compromised image quality in lower modulation orders despite the high symbol count. ME$_1$ exhibits better performance in low SNR regions, showcasing the strength of joint training. Nonetheless, as uJSCC mirrors TE's optimal performance, it delivers superior image quality in low SNR regions. This simulation confirms that uJSCC, with a robustly designed NN structure and training strategy, can be optimized to maintain satisfactory task quality across all SNR ranges when transmitting a higher number of symbols.

\subsection{uJSCC vs. Benchmarks with CelebA Dataset}

\begin{figure}[!t]
    \centering
    \subfigure[SSIM vs. SNR]{\includegraphics[width=0.85\linewidth]{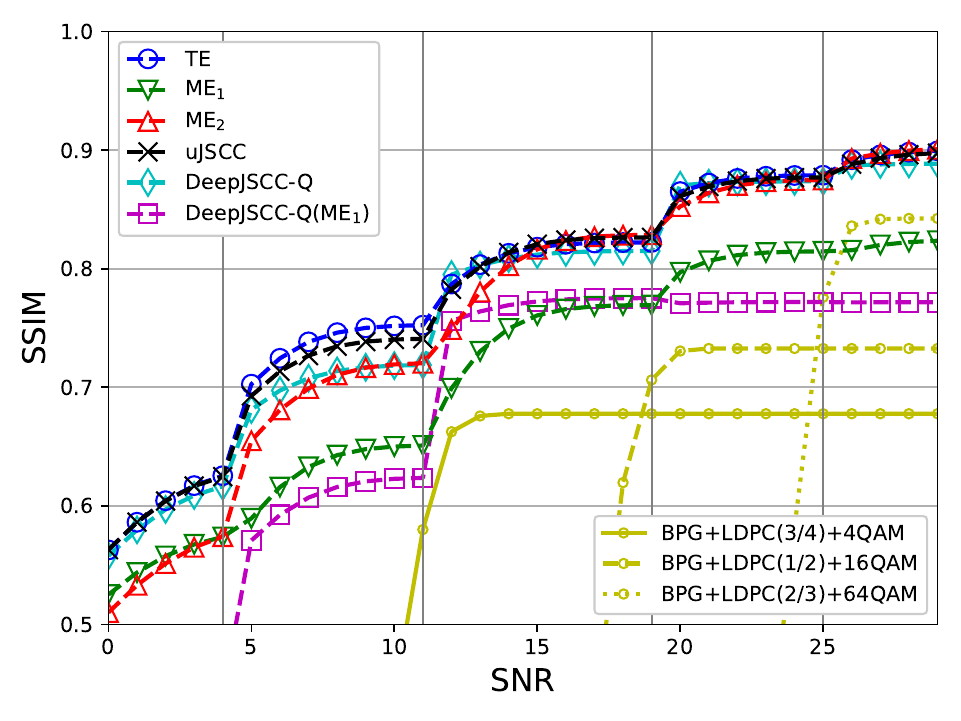}}
    \subfigure[PSNR vs. SNR]{\includegraphics[width=0.85\linewidth]{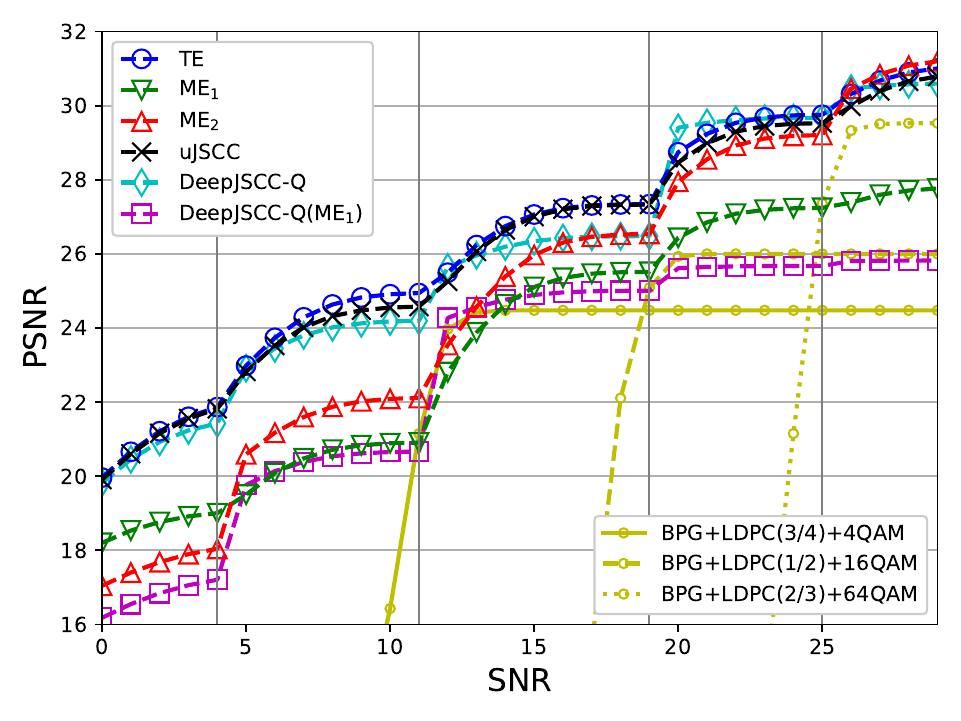}}
    \caption{Performance comparison between uJSCC vs. benchmark schemes using CelebA dataset.}
    \label{fig:Performance comparison with celeba dataset N=1024}
    \vspace{-5mm}
\end{figure}

In order to evaluate the proposed model with higher resolution images, we utilize the CelebA dataset \cite{liu2015faceattributes} cropped into the size of $3\times 128\times 128$. NN structure for \textbf{Large} setting is modified to generate $N = 1024$ symbols such that the second and the fourth layer of the outer encoder and decoder is changed to Conv($c_1, c_2, 4, 2, 1$) and T-Conv($c_2, c_1, 4, 2, 1$), respectively. When training uJSCC, $\alpha$ and $\lambda$ are set to ($2, 1.5, 1, 0.7, 0.5$) and ($1, 1, 2, 4, 16$), respectively. Hyperparameters for benchmark schemes are the same with \textbf{More Symbol} setting. For convenience, we refer to this configuration as \textbf{Basic} setting for the CelebA dataset.

\begin{figure}[!t]
    \centering
    \subfigure[SSIM vs. SNR]{\includegraphics[width=0.85\linewidth]{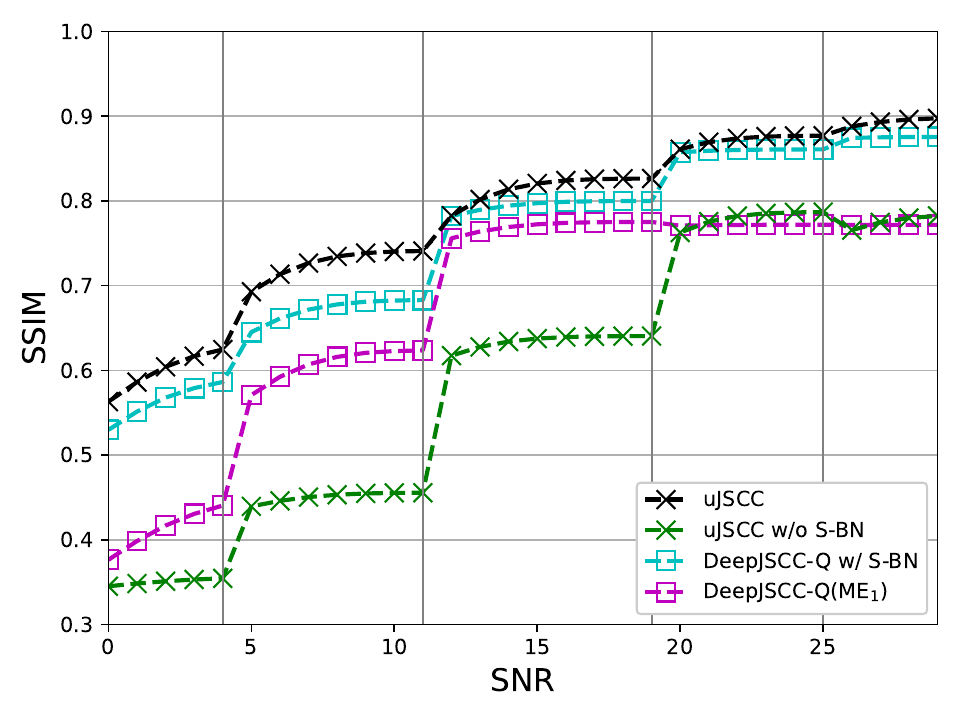}}
    \subfigure[PSNR vs. SNR]{\includegraphics[width=0.85\linewidth]{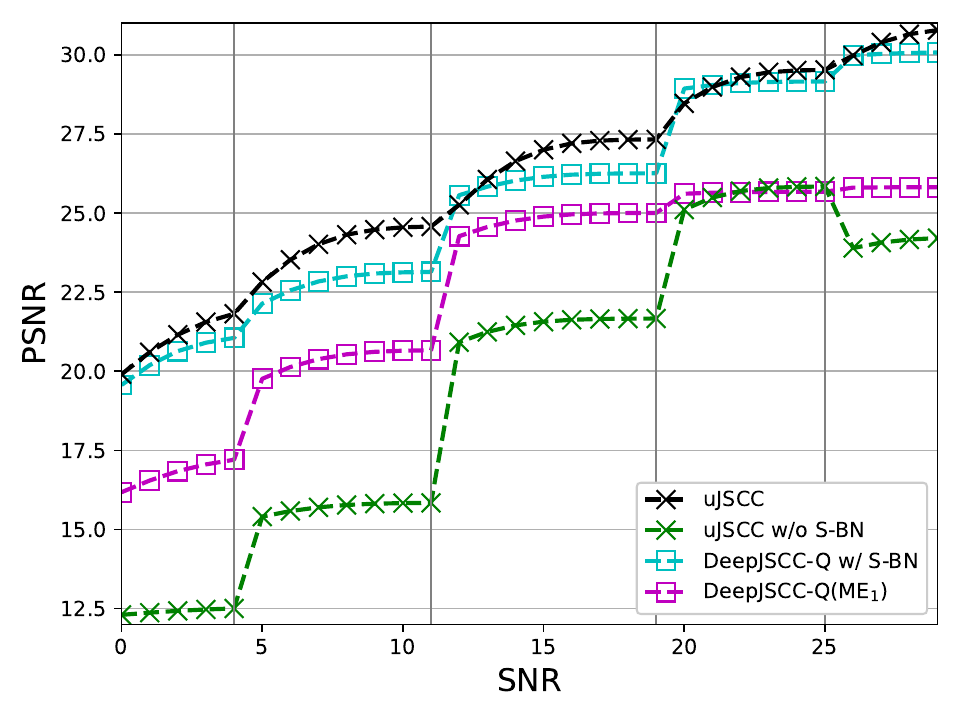}}
    \caption{Performance comparison between uJSCC vs. benchmark schemes selecting S-BN layer with CelebA dataset.}
    \label{fig:SBN ablation with celeba dataset N=1024}
    \vspace{-5mm}
\end{figure}

Fig. \ref{fig:Performance comparison with celeba dataset N=1024} verifies that uJSCC exhibits a consistent performance trend for high-resolution images. The proposed method closely aligns with the task performance of TE, showcasing that uJSCC, when combined with our training algorithm, effectively reaches the target capability. Additionally, uJSCC achieves higher SSIM and PSNR values for BPSK and 4QAM than DeepJSCC-Q, which is trained for a specific modulation order, and displays comparable performance across other modulation orders. Meanwhile, ME$_1$ continues to underperform across all SNR ranges, and ME$_2$ falls short of matching TE for BPSK, 4QAM, and 16QAM due to its training suboptimality. As uJSCC supports multiple modulation orders well with only a single encoder-decoder pair for lower-resolution CIFAR-10 images, it demonstrates similar adaptability and robust performance for high-resolution CelebA images. This result highlights the versatility and effectiveness of our proposed method.

\subsection{Ablation Study on S-BN Layer}

Fig. \ref{fig:SBN ablation with celeba dataset N=1024} illustrates the critical role of S-BN layers in enabling a single model to effectively and efficiently support multiple modulation orders with the CelebA dataset. uJSCC without S-BN layers exhibits significantly poor performance and even worse than DeepJSCC-Q in the ME$_1$ style, underscoring that S-BN layers are essential for uJSCC to achieve TE's performance. Furthermore, without S-BN layers, uJSCC struggles to show an improvement in the quality of reconstructed images as SNR increases into 256QAM range. Although DeepJSCC-Q performs below uJSCC considering both SSIM and PSNR values, DeepJSCC-Q with S-BN layers outperforms its ME$_1$ variant. These findings confirm that S-BN layers are crucial for a single neural network model with multiple data processing paths to enhance the task quality across each path by equalizing the data statistics.

\begin{figure}[!t]
    \centering
    \subfigure[CIFAR-10 dataset]{\includegraphics[width=0.9\linewidth]{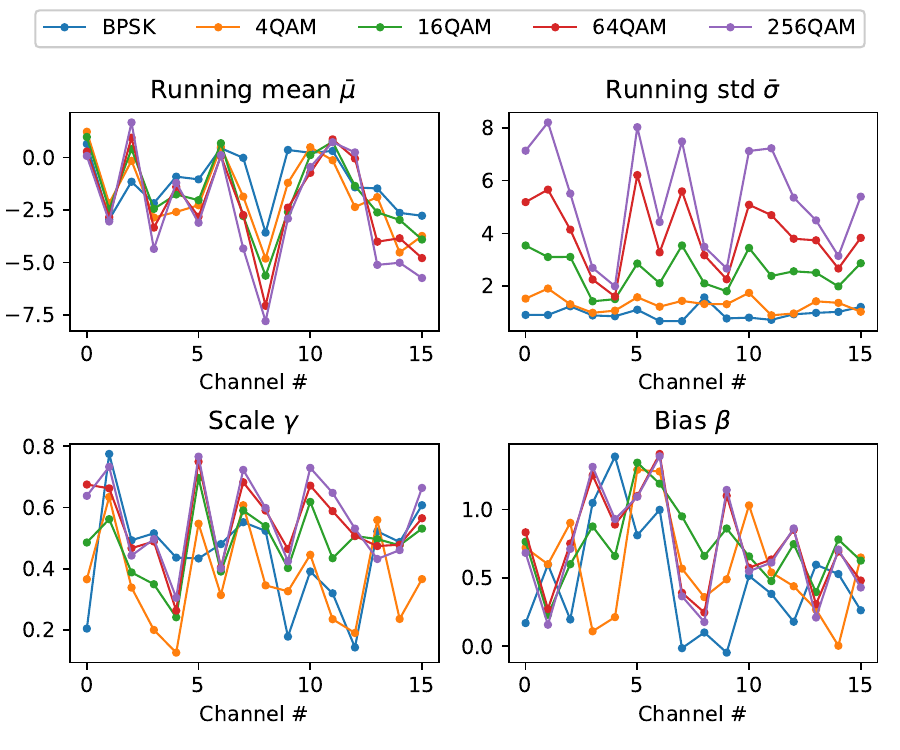}}
    \subfigure[CelebA dataset]{\includegraphics[width=0.9\linewidth]{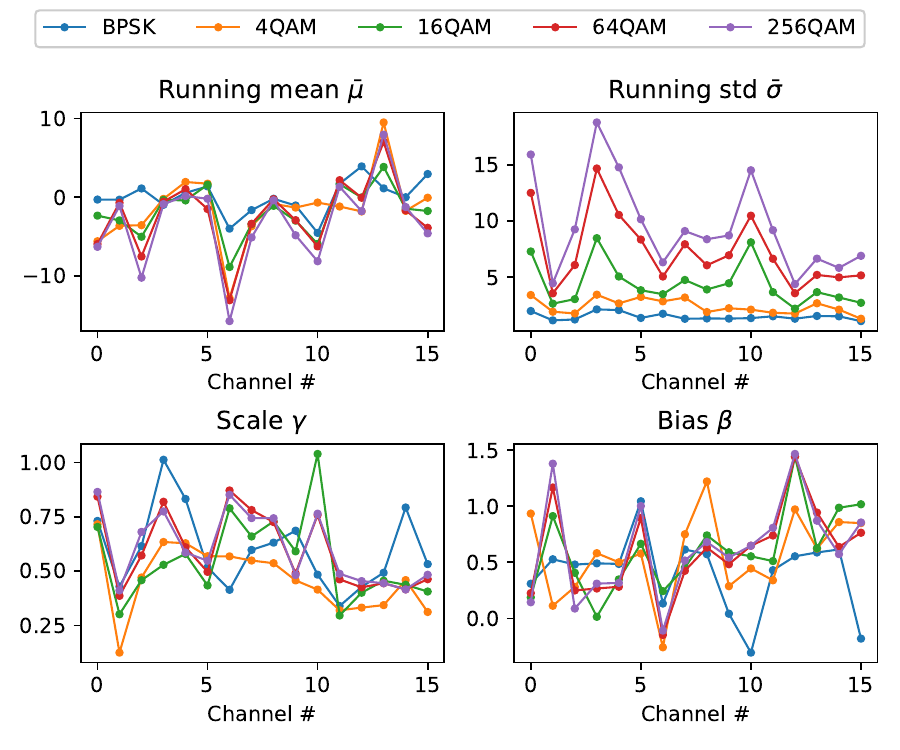}}
    \caption{Parameter values of the S-BN layer of uJSCC inner decoder under \textbf{Basic} setting.}
    \label{fig:SBN values}
    \vspace{-5mm}
\end{figure}

Fig. \ref{fig:SBN values} visualizes the actual values of parameters in the S-BN layer in each inner decoder for two datasets under \textbf{Basic} setting. Note that the input tensor size for the inner decoder $g_{\boldsymbol{\phi}_\mathrm{inner}}$ varies with each modulation order. Given that the decoder consists of T-Conv layer, the output tensor is generated by summing filtered tensors of the same size, with the summation repeated to match the number of channels in the input tensor. Consequently, as shown in the graphs of the running standard deviation, these values increase as the modulation order becomes higher. The fluctuation in the graphs also reflects that S-BN parameters used in forward operations vary across the five data processing paths. In particular, the learnable scale and bias parameters exhibit significant variation, highlighting the importance of S-BN layers to effectively support multiple modulation orders in uJSCC.

\subsection{Further Discussions for Future Work}\label{subsec:Further Discussion of Future Work}

While uJSCC achieves TE-level performance with notable efficiencies in model size, training, and data processing, there is room for improvement. We outline three key directions for future work below:
\begin{itemize}
\item \textbf{Learnable SNR boundaries}: Instead of intuitively setting SNR boundaries based on the 3GPP specification \cite{3gpp.38.214}, making these boundaries learnable could enhance performance by optimizing modulation order selection within SNR ranges, requiring updates to the training algorithm.
\item \textbf{Power-efficient codeword assignment}: The proposed algorithm lacks an average power constraint and does not regularize the uniformity of codeword usage. While uniform codeword usage balances power, it may impede JSCC learning. Instead, advanced methods could improve power efficiency by assigning high-power symbols to rare features without compromising task performance.
\item \textbf{SNR-adaptive feature generation}: Adaptive feature generation, inspired by \cite{xu2021wireless}, could be integrated into a VQ-based JSCC system by employing a hypernetwork \cite{ha2016hypernetworks} to fine-tune VQ codebooks for different SNRs, enabling optimal performance under varying channel conditions.
\end{itemize}

\section{Conclusion}\label{sec:Conclusion}
In this paper, we introduced a uJSCC framework that facilitates modulation-agnostic digital semantic communication. The data processing path within uJSCC mimics a traditional JSCC encoder-decoder pair tailored to a specific modulation order. Utilizing parameter sharing, which enhances model efficiency, the output statistics of each CNN layer are regulated by S-BN layers tailored for each modulation order to standardize the output statistics across different data processing paths. Through our proposed joint training strategy, experimental results on both low-resolution and high-resolution datasets demonstrate that uJSCC, with significantly fewer parameters and reduced training epochs, either matches or closely approaches the task performance of traditional encodings as evaluated by metrics such as SSIM, PSNR, and the quality of reconstructed images for each modulation order. Furthermore, we have validated the versatility of uJSCC in larger models and with different numbers of transmitted symbols. This study aims to illuminate the practical applicability of JSCC-based semantic communication using digital communication devices in real-world scenarios.

%
%% use section* for acknowledgment
% \section*{Acknowledgment}

\bibliographystyle{IEEEtran}  % appearance order
\bibliography{IEEEabrv,reference}

\end{document}